\documentclass[]{aa}
\usepackage{astro,natbib,graphicx,amssymb,amsmath,txfonts}
\bibpunct{(}{)}{;}{a}{}{,}

\begin{document}
\defcitealias{falgarone1998kp}{F98}
\defcitealias{hilyblant2007ii}{Paper~II}
\defcitealias{pety2003}{PF03}
\def \muvir{\ensuremath{\mu_{\rm vir}}}
\def \tpic{\ensuremath{T_{\rm peak}}}
\def \mcld{MCLD\,123.5+24.9}
\def \leq{\ensuremath{\Delta v_{\rm eq}}}
\def \re{\ensuremath{{\rm Re}}}
\def \vcar#1{\ensuremath{\overline{v}_{#1}}}
\def \vell{\ensuremath{v_{l}}}
\def \zz{\ensuremath{\tilde{\zeta}}}
\def \dc{\ensuremath{\delta C_l}}
\def \vecr{\ensuremath{\vec{r}}}
\def \vecl{\ensuremath{\vec{l}}}
\def \cpdf{\ensuremath{{\cal P}(\dc)}}
\def \abscpdf{\ensuremath{{\cal P}(|\dc|)}}
\def \absncpdf{\ensuremath{{\cal P}_n(|\dc|)}}
\def \ncpdf{\ensuremath{{\cal P}_n(\dc)}}
\def \sigdc{\ensuremath{\sigma_{\delta C}}}
\def \aver#1{\ensuremath{\langle #1 \rangle}}
\def \avcvi{\aver{|\dc|}} 
\def \sfe#1#2{\ensuremath{\epsilon_{#1}^{(#2)}}} 
\def \sff#1#2{\ensuremath{{\cal F}_{#1}^{(#2)}}}
\def \eps#1#2{\ensuremath{\aver{\epsilon_{#1}^{#2}}}}
\def \eeps#1#2{\ensuremath{\epsilon_{#1}^{(#2)}}}
\def \nmin{\ensuremath{N_{\rm min}}}
\def \flatness{\ensuremath{{\cal F}}}
\def \zzsl{\ensuremath{\zz_p^{\rm SL}}}
\def \zzmhd{\ensuremath{\zz_p^{\rm B02}}}
\def \zzik{\ensuremath{\zz_p^{\rm IK}}}
\def \shear{\ensuremath{\varpi}}
\def \visco{\ensuremath{\eta}}
\def \dissipm{\ensuremath{\epsilon}}
\def \dissipv{\ensuremath{\underline{\epsilon}}}
\def \dissip#1{\ensuremath{\epsilon_{#1}}}
\def \vth{\ensuremath{v_{\rm th}}}
\def \dv{\ensuremath{\langle \partial v \rangle}}
\def \lcow{\ensuremath{\Lambda_{\rm CO,wing}}}
\def \eturb{\ensuremath{\overline{\epsilon}_{\rm turb}}}
\def \efil{\ensuremath{\epsilon_{\rm fil}}}
\def \dfil{\ensuremath{d}}
\def \ldiss{\ensuremath{l_{\rm diss}}}
\def\sigaver{\ensuremath{\sigma_{\rm aver}}}


\title{Dissipative structures of diffuse molecular gas}
\subtitle{III -- Small-scale intermittency of intense
velocity-shears\thanks{Based on observations carried out
with the IRAM-30m telescope. IRAM is supported by
INSU-CNRS/MPG/IGN.}}

\author{%
  P.~Hily-Blant\inst{1,2}
  \and
  E.~Falgarone\inst{3}
  \and
  J.~Pety\inst{1,3}
}

\offprints{\email{philybla@obs.ujf-grenoble.fr}}

\institute{ IRAM, Domaine Universitaire, 300 rue de la
  Piscine, F-38406 Saint-Martin-d'H\`eres \and Laboratoire
  d'Astrophysique, Observatoire de Grenoble, BP 53, F-38041
  Grenoble Cedex 9\and LRA/LERMA, UMR 8112, CNRS,
  Observatoire de Paris and \'Ecole normale sup\'erieure, 24
  rue Lhomond, F-75231 Paris Cedex 05 }

\date{Received / Accepted}

\abstract {} {We further characterize the structures
tentatively identified on thermal and chemical grounds as
the sites of dissipation of turbulence in molecular clouds
(Papers I and II).}
{Our study is based on two-point statistics of line centroid
velocities (CV), computed from three large \twCO\ maps of
two fields.  We build the probability density functions
(PDF) of the CO line centroid velocity increments (CVI) over
lags varying by an order of magnitude.  Structure functions
of the line CV are computed up to the 6$^{th}$ order. We
compare these statistical properties in two translucent
parsec-scale fields embedded in different large-scale
environments, one far from virial balance and the other
virialized. We also address their scale dependence in the
former, more turbulent, field.}
{The statistical properties of the line CV bear the three
signatures of intermittency in a turbulent velocity field:
(1) the non-Gaussian tails in the CVI PDF grow as the lag
decreases, (2) the departure from Kolmogorov scaling of the
high-order structure functions is more pronounced in the
more turbulent field, (3) the positions contributing to the
CVI PDF tails delineate narrow filamentary structures
(thickness $\sim 0.02$ pc), uncorrelated to dense gas
structures and spatially coherent with thicker ones ($\sim
0.18$ pc) observed on larger scales.  We show that the
largest CVI trace sharp variations of the extreme CO
linewings and that they actually capture properties of the
underlying velocity field, uncontaminated by density
fluctuations. The confrontation with theoretical predictions
leads us to identify these small-scale filamentary
structures with extrema of velocity-shears. We estimate that
viscous dissipation at the 0.02~pc-scale \textit{in these
structures} is up to 10 times higher than average,
consistent with their being associated with gas warmer than
the bulk.  Last, their average direction is parallel (or
close) to that of the local magnetic field projection.  }
{Turbulence in these translucent fields exhibits the
statistical and structural signatures of small-scale and
inertial-range intermittency.  The more turbulent field on
the 30~pc-scale is also the more intermittent on small
scales. The small-scale intermittent structures coincide
with those formerly identified as sites of enhanced
dissipation. They are organized into parsec-scale coherent
structures, coupling a broad range of scales. }

 \authorrunning{Hily-Blant
  P. \etal} \titlerunning{Intermittency in diffuse molecular
  gas} \maketitle

\def\vlos{\ensuremath{v_{\rm los}}}

\section{Introduction}

Star formation proceeds \via\ gravitational instability in
dense gas, but the respective roles of turbulence and
magnetic fields in that process are still debated issues
\citep{ciolek2006, tassis2004, maclow2004, padoan2001,
klessen2001, bate2002} in spite of dedicated observational
studies of magnetic fields in molecular clouds
\citep{matthews2000,crutcher1999} and extensive theoretical
and numerical works devoted to characterizing the properties
of the turbulence \citep{boldyrev2002,padoan2003}. A hybrid
paradigm is taking shape, where turbulence dominates the
diffuse ISM dynamics and magnetic fields gain importance as
the scale decreases \citep{crutcher2005}. Turbulence and
magnetic fields are recognized as powerful stabilizing
agents in molecular clouds, and a critical issue remains,
turbulence dissipation: where, when, and at which rate and
scale does it occur~?

A generic property of turbulent flows is the space-time
intermittency of the velocity field.  Intermittency is
observed in laboratory experiments of incompressible
turbulence, in the atmosphere, and in the solar wind
\citep[see recent reviews by ][]{anselmet2001,bruno2005}.
It manifests itself \textit{i)} as non-Gaussian tails in the
probability distribution of all quantities involving
velocity differences (\eg\ gradient, shear, vorticity, rates
of strain, and energy dissipation rate), \textit{ii)}
anomalous scaling of the high-order structure functions of
the velocity increments \citep{anselmet1984}, and
\textit{iii)} the existence of coherent structures of
intense vorticity \citep[][ hereafter
MJ04]{vincent1991,moisy2004}. It has long been unclear
whether these three manifestations, which refer either to
the statistical properties of the flow or to its coherent
structures, were related.

Statistical models make theoretical predictions in terms of
the two-point statistics of the velocity field, with no link
to any physical structures in the turbulent flow. In
particular, the structure functions $S_p(l)=\langle
[v(r+l)-v(r)]^p\rangle$ of the velocity field are expected
to be power laws $S_p(l) \propto l^{\zeta(p)}$. As $p$
increases, the structure functions give more weight to
intense and rare events.  In principle then, a detailed
description of the velocity field could be achieved with the
knowledge of all $S_p$ for $p\rightarrow\infty$, however,
the number of points $N_p$ needed to compute $S_p$ grows
with $p$, and an educated guess is $N_p \sim
10^{p/2}$. Fortunately, theoretical models show that, with a
limited number of orders ($p>3$), some properties of the
turbulence can still be tested. The Kolmogorov theory of
non-intermittent turbulence \citep[hereafter K41, see
\eg][]{frisch1995} predicts $\zeta(p)=p/3$, while
experiments show a clear departure from this scaling,
generally with $\zeta(p)<p/3$ for $p>3$. The departure from
the $p/3$ scaling is usually interpreted as a definition of
intermittency. Statistical models predict intermittent
scalings $\zeta(p)\ne p/3$. One such model \citep[][
hereafter SL94, see Appendix~A3]{she1994} has an
intermittent scaling $\zzsl = p/9 + 2[ 1-(2/3)^{p/3}]$, in
excellent agreement with tunnel-flow experimental data
\citep{benzi1993}.

The structural approach is inspired from laboratory
experiments showing filaments of high vorticity
\citep{douady1991}. Localized regions of high vorticity and
rate of strain (and thus energy dissipation) are found in
numerical simulations at high resolution of both
incompressible and compressible \citep{porter1994} and
magneto-hydrodynamical (MHD) turbulence
\citep{vincent1991,mininni2006b}. Recently, MJ04 found that
intense structures of vorticity and rate of strain are
respectively filaments and ribbons that are not randomly
distributed in space but that instead form clusters of
inertial-range extent, implying a large-scale organization
of the small-scale intermittent structures.  In $1024^3$
numerical simulations of incompressible turbulence, with
variable large-scale stirring forces, \cite{mininni2006a}
have shown that the large and small-scale properties of the
flow are correlated, namely that \textit{i)} more intense
small-scale gradients and vortex tubes are concentrated in
the regions where the large-scale shears are the largest,
and that \textit{ii)} the departure from the Kolmogorov
scaling is more pronounced in these regions. They infer from
these results that the statistical signatures of
intermittency are linked to the existence of the small-scale
vortex tubes.

\begin{figure}[t]
  \begin{center}
    \includegraphics[width=0.35\textwidth]{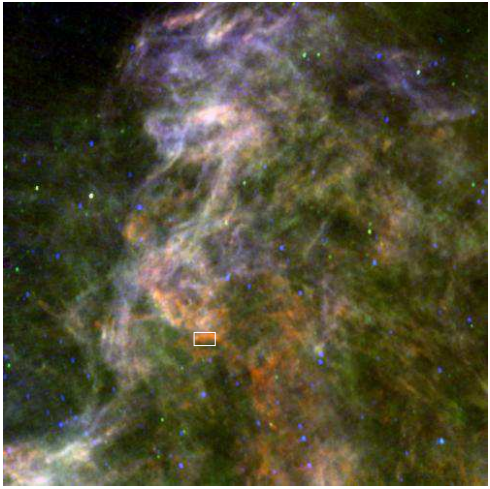}
    \caption{Reprocessed IRAS map of the Polaris Flare
      \citep{miville2005}. The map size is about 10$\degr
      \times 10\degr$ or $27\times27$~pc. The parsec-scale
      field analyzed in this paper is shown as a white
      rectangle. The 100$\mu$m, 60$\mu$m and 12$\mu$m
      emissions are red, green and blue respectively.}
    \label{fig:polaris-iris}
  \end{center}
\end{figure}

\begin{figure}[t]
  \begin{center}
    \includegraphics[width=\hsize,angle=0]{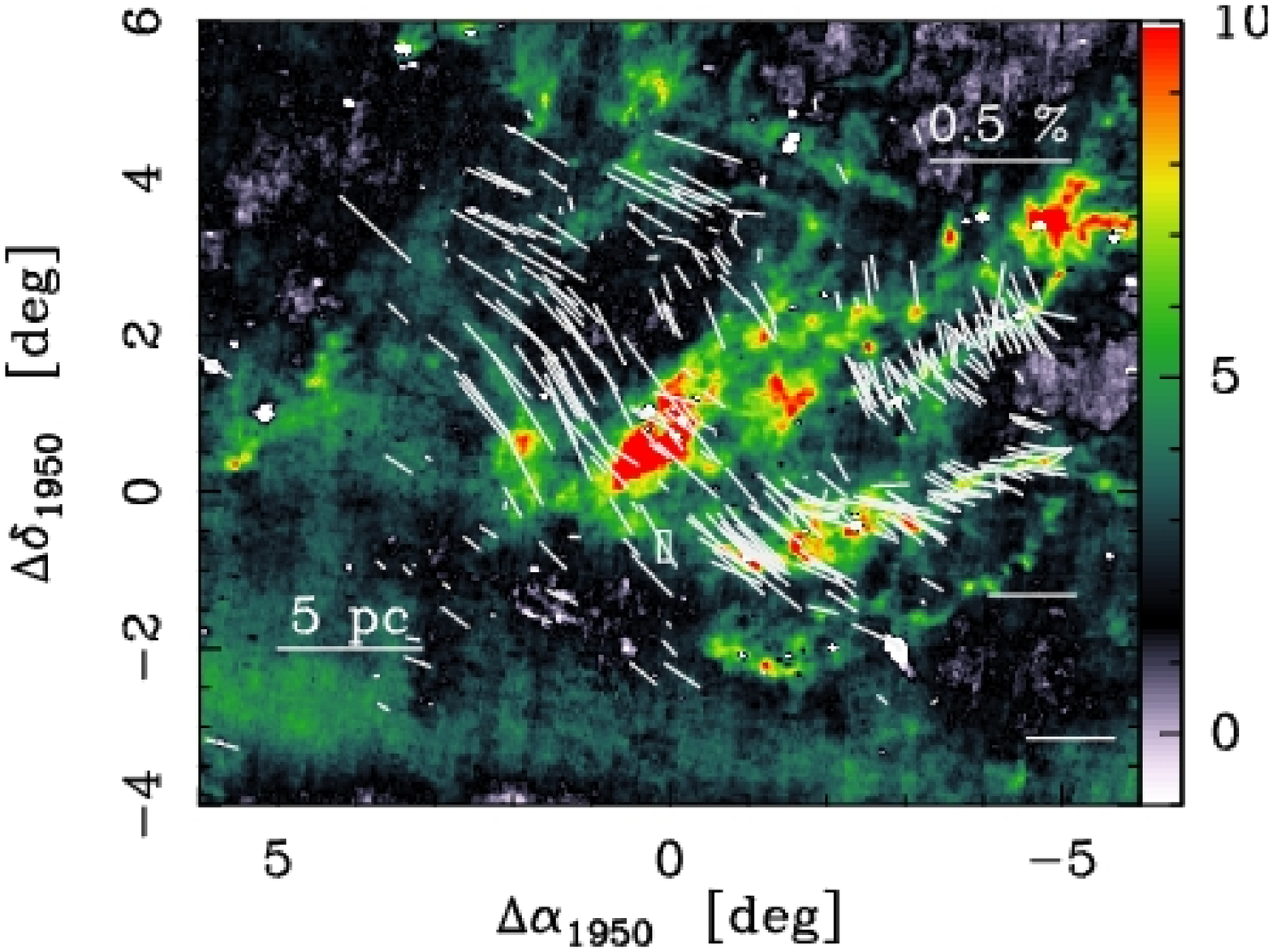}
    \caption{Map of the cold dust emission in the Taurus
      molecular complex built from the reprocessed 60$\mu$m
      and 100$\mu$m IRAS maps \citep{hilyblant2007sins}
      following the method of \cite{abergel1995}. The
      orientation of the projection of the magnetic fields
      on the sky \citep{heiles2000} is shown as white
      segments of length proportional to the polarization
      degree. The parsec-scale field discussed in this paper
      is centered at $\alpha_{2000}=\hms{04}{40}{08.84},
      \delta_{2000}=\dms{24}{12}{48.40}$ and is shown as a
      white rectangle at offsets (0\degr,-1\degr).}
    \label{fig:taurus-iris}
  \end{center}
\end{figure}

Investigations of interstellar turbulence are plagued by
projection effects that affect our knowledge of the
velocity. Direct inversion of the observations being a
highly degenerate procedure, progress relies on
astrophysical observables derived from numerical simulations
of turbulence, and their confrontation to real data
\citep[see the review of ][]{elmegreen2004}. Along these
lines, \cite{lis1996} have shown that it is possible to
trace the projection of the most intense velocity-shears
with a measurable quantity based on two-point statistics of
the velocity field: the line centroid velocity increments
(CVI). This method, applied to different fields observed in
CO lines, a star-forming region \citep{lis1998b} and
quiescent regions \citep[][ hereafter PF03]{pety2003}
revealed filamentary structures unrelated with the gas mass
distribution. The line centroid velocities (and their
increments) are sensitive to optical depth effects and to
density, temperature, and abundance fluctuations along the
line of sight. Their relevance in any analysis of the
statistical properties of the actual velocity fields have
therefore been repeatedly questioned. Several recent studies
have clarified the issue
\citep{lazarian2000,miville2003b,levrier2004,ossenkopf2006}
without having provided any final answer to that question
yet.  The present work is part of a broad study in which we
characterize the structures of largest CVI on thermal,
chemical, and statistical grounds, and then repeat this
analysis in different turbulent clouds. The goal of this
broad program is to shed light on what these structures
actually trace.

One of the two fields studied in this paper is the
parsec-scale environment of two low-mass dense cores in the
Polaris Flare. In the vicinity of the dense cores,
\cite{falgarone2006hcop} (hereafter Paper~I) find \hcop\
abundances locally far in excess of steady-state chemical
predictions.  The large measured abundances are consistent
with a scenario that involves an impulsive heating of the
gas, during which a warm chemistry is triggered, followed by
a slow chemical and thermal relaxation. According to this
scenario, the observed abundances and gas densities, in the
range 200 and 600~\ccc, correspond to a gas that has cooled
down to 100--200~K.  In \cite{hilyblant2007ii} (hereafter
Paper~II), we analyze the mass distribution of the gas in
the environment of the dense cores and disclose localized
regions where the \twCO\ profiles exhibit broad
wings. Multi-line analysis shows that gas in these linewings
is optically thin in the \twCO\jone\ line and warmer than
25~K with density $<1000$\,\ccc. The new result is that this
warm gas component is not widespread but localized in
filaments.  The dispersion of the orientation of the
filaments of dense gas in the field, compared to the
velocity dispersion, suggests that the turbulence in that
field is trans-Alfv\'enic. In the forthcoming and last paper
\citep{falgarone_pdbi}, we will report \twCO\jone\
IRAM-Plateau de Bure observations performed in this field,
revealing milliparsec-scale structures with large
velocity-shears.
 
The present paper is dedicated to the statistical and
structural analysis of turbulence in similar parsec-scale
translucent samples of gas, which belong to two different
large-scale environments: one, far from virial balance and
devoid of stars, is the above-mentioned field in the Polaris
Flare; the other lies at the edge of the Taurus-Auriga
molecular complex.  The statistical analysis is based on
two-point statistics of the CO line emission observed at
high spectral resolution, and the structural analysis
consists in characterizing the regions of largest
line-centroid velocity increments.

After presenting the data in Section~2, the characteristics
of the turbulence in these two fields are derived based on
the PDF of line CVI and structure functions (Section~3). We
then show, in Section~4, that the regions of largest CVI on
small-scales are filaments uncorrelated with the
distribution of matter, but correlated with the filaments of
gas optically thin in \twCO\jone\ where large HCO$^+$
abundances are found. It is shown that these filaments
remain coherent at the parsec-scale.  In Section~5, we show
that the ensemble of results, based either on the
statistical or structural approach, provides a consistent
description of the intermittency of turbulence in these two
fields.  In Section~6, the comparison of the radiative
cooling of these structures with the fraction of the
turbulent energy susceptible to being dissipated there
further supports the proposition that the filaments of
largest CVI somehow trace extrema of velocity-shear in the
fields and pinpoint the sites of intermittent dissipation of
turbulence.

\section{Observations}

\begin{figure}
  \begin{center}
    \includegraphics[width=0.7\hsize]{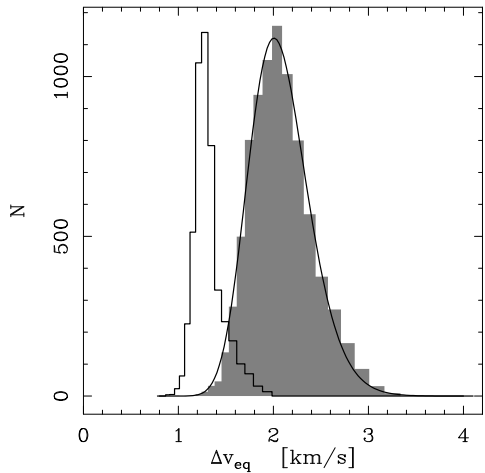}
    \caption{Histograms of the equivalent width ($\leq=\int
      T(v_x) \ud v_x/\tpic$) of the \twCO\jone\ line in
      Polaris (shaded area) and Taurus. The result of the
      log-normal fit to the Polaris histogram is also
      shown.}
    \label{fig:dveq}
  \end{center}
\end{figure}

\subsection{Polaris field}
  
Paper~II describes the observations and data reduction of
the IRAM \twCO\ and \thCO\jone\ and \jtwo\ maps of the
molecular cloud \mcld\ located in the Polaris Flare. The
location of the field mapped is shown in
Fig.~\ref{fig:polaris-iris}. The fully-sampled
$\approx1\,\pc\times\pc$ maps cover the non--star-forming
translucent environment of two dense cores
\citep{gerin1997,falgarone1998kp,heithausen2002},
corresponding to $\approx 3000$ independent spectra (\ie\
spaced by one beamsize or 20\arcsec\ at 115~GHz). The
spectral resolution is 0.055\kms. Since the signal-to-noise
ratio of the $J=2-1$ data is insufficient, only the $J=1-0$
transition are used in the present paper. For comparison
with larger scale properties, we use the fully-sampled
\twCO\jtwo\ data from \cite{bensch2001} obtained at a lower
angular resolution ($HPBW=120\arcsec$) with the 3m KOSMA
antenna.

\subsection{Taurus field}

The second field is located at the edge of the Taurus-Auriga
molecular complex (Fig.~\ref{fig:taurus-iris}).
Observations were done with the IRAM-30m
telescope. Observational strategy and data reduction are
similar to those of the Polaris field. The maps, centered at
($\alpha_{2000}=\hms{04}{40}{08.84},
\delta_{2000}=\dms{24}{12}{48.40}$), are fully sampled in
\twCO\ and \thCO\jone. The data will be presented in more
detail in a later paper \citep{hilyblant_taurus} but here we
give the properties relevant to the present work. A total of
1200 independent spectra was obtained with the same spectral
resolution of 0.055\,\kms. In a small region (around
0,0\arcsec\ offsets in the map of
Fig.~\ref{fig:cvimap-lag3}), spectra show two separate
components (at $v_{\rm LSR}\approx5$ and 10~\kms), and the
analysis presented in this paper focuses on the main
component at $v_{\rm LSR}\approx5$~\kms, by blanking out the
area corresponding to the high-velocity component.

\subsection{Comparison of the two fields}

\begin{table}
  \begin{center}
    \caption{Dispersions \sigdc\ (in \kms) of the PDF of CVI
      computed from the \twCO\jone\ transition in the
      Polaris and Taurus fields for different lags $l$ (in
      pixels).}
    \begin{tabular}{c c c c }\hline\hline
      \multicolumn{2}{c}{$l$} & \multicolumn{2}{c}{\sigdc} \\
      & & Polaris &  Taurus \\
      $[$pixels$]$ & [pc] & \kms & \kms \\\hline
       3      &   0.02   & 0.11  & 0.05 \\
       6      &   0.05   & 0.17  & 0.07 \\
       9      &   0.07   & 0.22  & 0.09 \\
      12      &   0.09   & 0.25  & 0.10 \\
      15      &   0.11   & 0.27  & 0.11 \\
      18      &   0.14   & 0.29  & 0.12 \\
      18$^{(a)}$  &   0.14   & 0.30  &  --  \\
      36$^{(a)}$  &   0.27   & 0.46  &  --  \\
      54$^{(a)}$  &   0.41   & 0.57  &  --  \\
      72$^{(a)}$  &   0.54   & 0.64  &  --  \\
      90$^{(a)}$  &   0.68   & 0.69  &  --  \\
     108$^{(a)}$  &   0.81   & 0.72  &  --  \\\hline
    \end{tabular}
	\begin{list}{}{}
	  \scriptsize
	\item[$(a)$] Computed from the KOSMA \twCO\jtwo\ data of
      \cite{bensch2001}, where one pixel corresponds to 6
      pixels of the IRAM data. The adopted distance is
      150~pc for both fields.
	\end{list}
	\label{tab:sig}
  \end{center}
\end{table}

Both fields are translucent. The visual extinction lies
between $\av=0.6$ and 0.8~mag at a resolution of 8\arcmin\
in the Polaris field \citep{cambresy2001} and between
$\av=1$ and 1.2~mag in the Taurus field at the same
resolution \citep{cambresy1999}.  As shown in
Fig.~\ref{fig:dveq}, the most probable equivalent width
($\leq=\int T(v_x) \ud v_x/\tpic$, $x$ being the coordinate
along the line of sight) of the \twCO\jone\ line is a factor
two larger in Polaris than in Taurus. However, since the
lines are stronger in the Taurus field, the integrated
intensities in both fields are similar. Not only is the
equivalent width larger in the Polaris field than in the
Taurus one, but so is the dispersion of these equivalent
widths. This factor 2 between the equivalent width
translates into a factor 4 in the specific kinetic energy
ratio between the two fields. The distribution of the
equivalent width in Polaris is also very well-fitted by a
log-normal distribution centered at $\leq = 2$~\kms\ with
dispersion 1.2~\kms. The parsec-scale velocity gradients,
deduced from the centroid velocity maps, in the Polaris
field ($\approx 2$~\kmspc) is also twice larger than in the
Taurus field.

These two parsec-scale fields are similar with respect to
their size and average column density but have specific
kinetic energies that differ by a factor 4.  Moreover,
they belong to two very different environments on the scale
of $\sim$ 30 pc, that of Figs.~\ref{fig:polaris-iris} and
\ref{fig:taurus-iris}.  The total gas mass
$M_{tot}=4.4\times 10^4$~\msol\ at the scale of 30~pc is
close to the virial mass in the Taurus-Auriga field
\citep{ungerechts1987}, while it is more than six times
lower in the Polaris Flare, $M_{tot}=5500$~\msol\ with $M_v=
3.6\times 10^4$~\msol\ \citep{heithausen1990}.  It is
interesting that the virial masses of the two large-scale
fields are close, because their velocity dispersion on the
scale of 30~pc are similar, 3.8 and 4.8~\kms, respectively.

In summary, the less turbulent parsec-scale field lies on
the far outskirts of the virialized Taurus-Auriga molecular
complex, while the more turbulent field belongs to a much
less massive complex, far from virial balance.


\section{Two-point statistics of the centroid velocity}


\subsection{Probability density functions of the line centroid velocity increments}

\begin{figure}[t]
  \begin{center}
    \includegraphics[width=0.65\hsize,angle=-90]{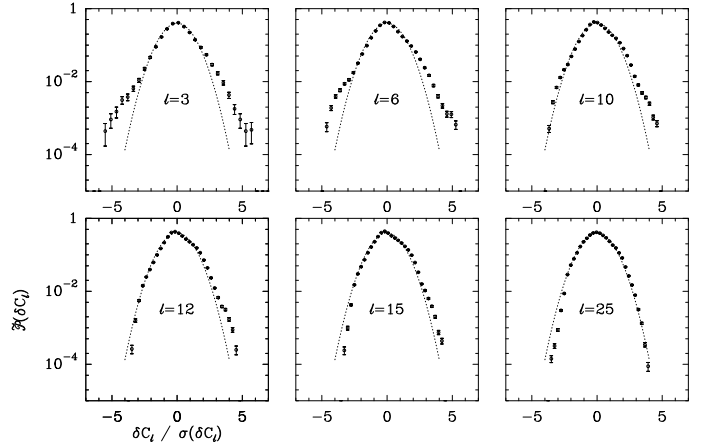}
    \caption{Normalized PDF, \ncpdf, of the centroid
      velocity increments (CVI) computed from the
      \twCO\jone\ IRAM map in Polaris . The PDF are computed
      for different lags between pairs of points: $l = 3$ to
      25 pixels, and normalized to unity dispersion, such
      that the $x-$axis is in units of the rms $\sigdc$ for
      each distribution. Only the bins containing more than
      10 data points were kept. The values of the dispersion
      $\sigdc$ are given in Table~\ref{tab:sig}. A Gaussian
      of unit dispersion is also shown (dotted curve).}
    \label{fig:pdf-polaris}
  \end{center}
\end{figure}

Following \cite{lis1996}, we analyze the two-point
statistics of the centroid of the line-of-sight projection
of the velocity $v_x$, which we note as $C(y,z)=C(\vecr)$,
where $(y,z)$ is the position on the sky:
\begin{equation}
  C(\vecr)=\int T(\vecr,v_x) v_x \ud v_x / \int T(\vecr,v_x)\ud v_x.
  \label{eq:c}
\end{equation}
Increments of the centroid velocity between 2 points
separated by $\vecl$ are defined by $\delta
C(\vecr,\vecl)=C(\vecr+\vecl)-C(\vecr)$. This quantity will
be called centroid velocity increment (CVI). The main
difficulty of this method concerns the computation of $C$,
which not only depends on the bounds of the integrals in
Eq.~\ref{eq:c}, but is also affected by the signal-to-noise
ratio (SNR). To circumvent the bias introduced by spatial
noise variations, \cite{rosolowsky1999} degrade all the
spectra to a unique threshold SNR. A different approach
(PF03) has been adopted here, which uses the SNR of the
integrated area as the optimization criterion to determine
the spectral window used to compute $C$. The reason is that
we have checked that the noise is already homogeneous in the
data cubes, mostly as a result of the observing strategy
consisting in several coverages of individual sub-maps in
perpendicular directions.

\begin{figure}[t]
  \begin{center}
    \includegraphics[width=0.65\hsize,angle=-90]{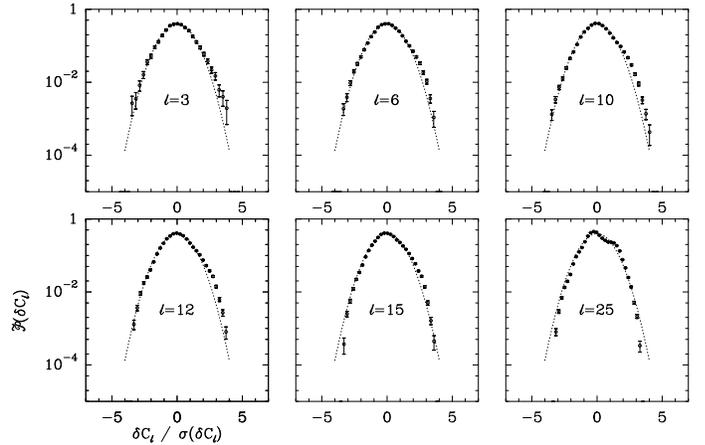}
    \caption{Same as Fig.~\ref{fig:pdf-polaris} for the
      \twCO\jone\ data towards the Taurus field.}
    \label{fig:pdf-taurus}
  \end{center}
\end{figure}

For a given value of $l=|\vecl|$, we compute maps of CVI for
each direction $\vecl/l$. A probability density function
(PDF) is built from these maps by normalizing the histogram
of CVI to a unit area. We thus obtain a PDF for each $l$ and
each direction. For each $l$, a PDF is computed with the CVI
from all directions $\vecl/l$, which we denote as
$\cpdf$. In order to get PDF with zero average and unit
standard deviation and to ease the comparisons, we use the
normalized PDF \ncpdf.  All bins of the \ncpdf\ associated
to a number of points less than a given value \nmin\ are
blanked (see Appendix A.1). In the following, all \ncpdf\
have 32 bins, and the adopted minimum number of data points
for a bin to be significant is $\nmin=10$. In a second step,
for a given $l$, we compute the azimuthal average of the
absolute value of the CVI, resulting in a single CVI map. In
practice the structures seen in the non-averaged maps are
not smeared out, though they appear thinner in some cases.

Figs.~\ref{fig:pdf-polaris} and \ref{fig:pdf-taurus} show
the \ncpdf\ computed for various lags from $l=3$ to 25
pixels in the Polaris and Taurus fields, respectively. The
lag $l=3$ is the shortest distance between two independent
points (since the sampling is half the beam size), and
$l=25$ corresponds to the largest lag with significant
number of pairs of points. The number of data points
corresponding to the three most extreme bins for $l=3$ and
25 are in the range $12-50$ and $20-500$, respectively.

\begin{figure}[t]
  \begin{center}
    \includegraphics[width=0.65\hsize,angle=-90]{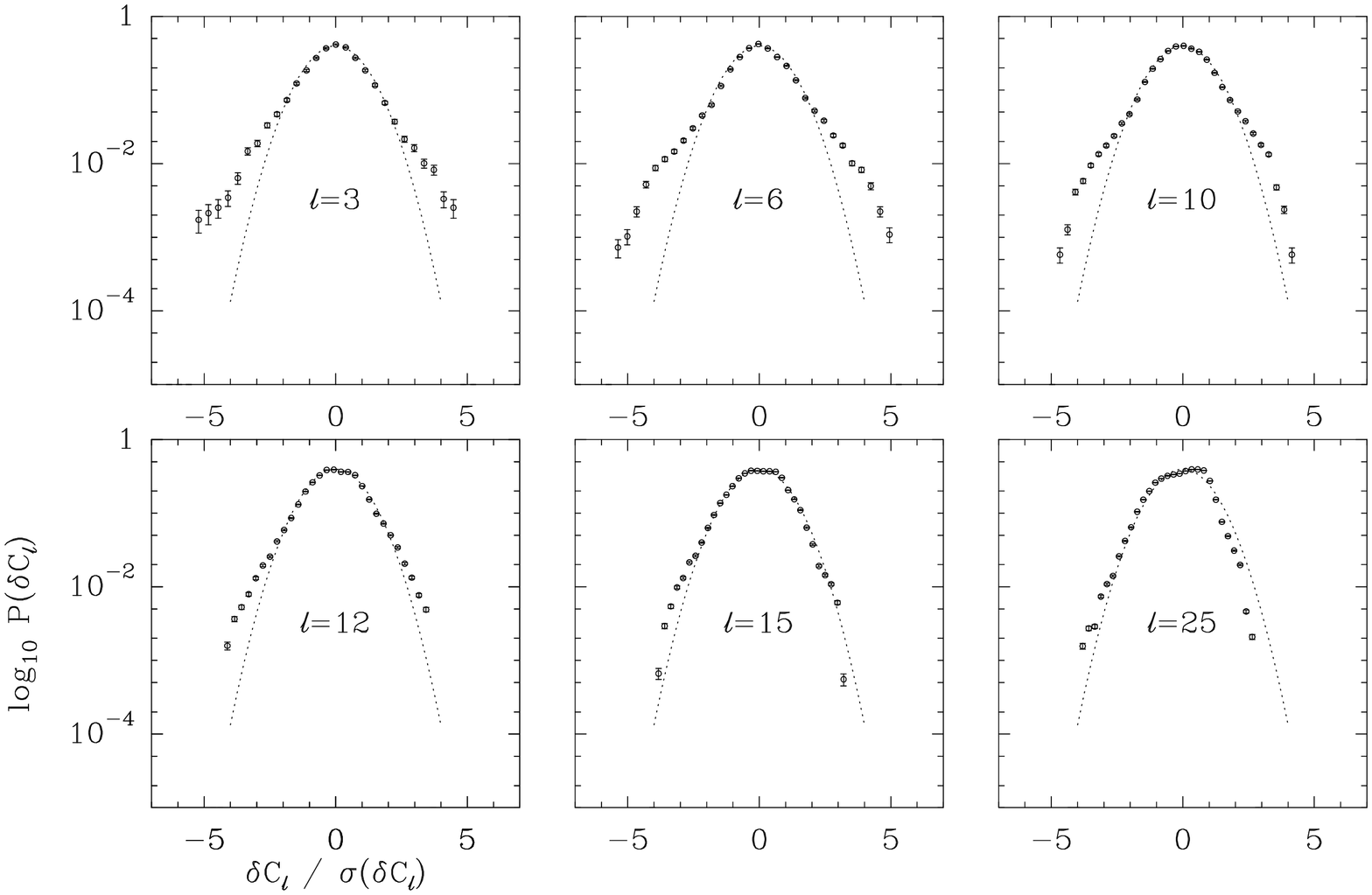}
    \caption{Same as Fig.~\ref{fig:pdf-polaris} for the
      \twCO\jtwo\ KOSMA data from \cite{bensch2001}.}
    \label{fig:pdf-kosma}
  \end{center}
\end{figure}

The PDF at large lags ($l>15$) (Fig.~\ref{fig:pdf-polaris})
are nearly Gaussian and become slightly asymmetrical at
$l=12$, an effect we attribute to large-scale velocity
gradients. Such effects cancel out at lags smaller than the
characteristic scale of these gradients, hence the more
symmetrical shape of the PDF at small lags. It is not
obvious that such large-scale gradients should be removed
(see discussion in PF03). The fields mapped here are
expected to be small with respect to the integral scale of
turbulence $L$, at least on the order of the molecular cloud
size itself. As the lag decreases from $l=25$ to $l=3$
pixels, non-Gaussian tails develop. These tails are more
pronounced in the Polaris field than in the Taurus one, with
CVI values up to 6 times the dispersions \sigdc\ of the
unnormalized PDF (see Table~\ref{tab:sig}). However, since
the number of points in Taurus is lower than in the Polaris
field, the minimum level of probability reached is an order
of magnitude higher (\dix{-3} instead of \dix{-4} in
Polaris). We also computed the PDF of the increments for the
large-scale KOSMA data in the Polaris field
(Fig.~\ref{fig:pdf-kosma}) and we also find increasing
non-Gaussian tails as the lag decreases. The dispersions of
the PDF are reported in Table~\ref{tab:sig} and are seen to
smoothly connect with the small-scale values computed in the
IRAM field.

In both fields, the dependence of \sigdc\ with $l$ can be
well-fitted by a power law $\sigdc\propto l^{0.5}$. The
dispersions \sigdc\ in the Taurus field are a factor
$\approx2$ smaller than in the Polaris field. This ratio is
also found when comparing the velocity dispersions -- either
across the plane of the sky ($pos$) or along the line of
sight ($los$) -- in the two fields (see
Table~\ref{tab:3dvel}).  Furthermore, the ratio of the $los$
to $pos$ dispersions suggests that the depth of the cloud is
larger than the extension in the plane of the sky
($los>pos$) \citep{ossenkopf2002}.

\subsection{Non-Gaussianity: the flatness}

\begin{figure}[t]
  \begin{center}
    \includegraphics[width=0.75\hsize]{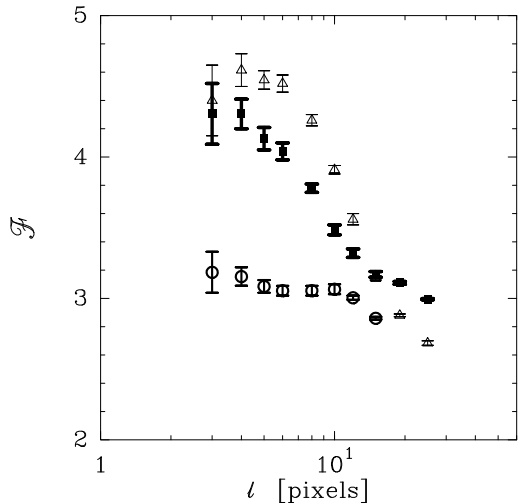}
    \caption{Flatness \flatness\ of the CVI (see
      Eq.~\ref{eq:flat}) computed from the normalized PDF of
      CVI. \emph{Squares}: Polaris \twCO\jone\ IRAM
      data. \emph{Triangles}: \twCO\jtwo\ KOSMA
      data. \emph{Circles}: Taurus \twCO\jone\ IRAM data.}
    \label{fig:flat}
  \end{center}
\end{figure}

The deviations of the PDF from a Gaussian shape can be
quantified using the flatness (or kurtosis) of a
distribution, defined by
\begin{equation}
  \flatness(l) = \frac{\langle {\dc}^4 \rangle}{\langle
  {\dc}^2 \rangle^2}
  \label{eq:flat}
\end{equation}
where the $p$th-order moment (for even $p$) is computed as
$\aver{\dc^p}=\int \dc^p\,\ncpdf\,\ud(\dc)$. The flatness
equals 3 for a Gaussian distribution. The uncertainties,
here and for all the moments computed in what follows, are
estimated by using two \ncpdf\ with two thresholds,
$\nmin=1$ and 10, and by computing the mean and rms of the
two outputs. Fig.~\ref{fig:flat} displays the flatness of
the PDF at all lags, for the two fields. For the Polaris
field, the IRAM and KOSMA data have a flatness close to 3 at
large lags, which confirms the visual Gaussian shape of the
corresponding PDF in Fig.~\ref{fig:pdf-polaris}. The
flatness increases at smaller lags as a result of
non-Gaussian tails. For the Taurus field, however, the
flatness stays close to 3, confirming that the non-Gaussian
tails are less pronounced.

\begin{table}
  \begin{center}
    \caption{Standard deviations $\sigma$ (in \kms) of the
      centroid velocity PDF ($pos$) and of the average line
      profiles ($los$), computed in three ways in the two
      fields ($\sigma_1$, $\sigma_2$, $\sigma_3$).}
    \begin{tabular}{c c c c c c c} \hline \hline
      Field & Type & Size & $\sigma_1$ & $\sigma_2$ & $\sigma_3$ & \aver{\sigma}\\
      (1) & (2)  & (3) & (4) & (5) & (6) & (7) \\ \hline
      Polaris$^\dag$  &$pos$ &  $2.1\times2.8$  & 0.54   & 0.57 & 0.60 & $0.57\pm0.03$\\
      &$los$ &                  & 1.13   & 1.10 & 1.16 & $1.13\pm0.03$ \\
      Polaris$^\ddag$ &$pos$ &  $0.7\times0.6$  & 0.31   & 0.32 & 0.25 & $0.29\pm0.04$ \\
      &$los$ &                  & 0.97   & 1.02 & 1.10 & $1.03\pm0.05$ \\
      Taurus$^\ddag$  &$pos$ &  $0.4\times0.7$  & 0.12   & 0.13 & 0.11 & $0.12\pm0.01$ \\
      &$los$ &                  & 0.50   & 0.60 & 0.66 & $0.59\pm0.07$ \\
      \hline
    \end{tabular}
    \label{tab:3dvel}
  \end{center}
  \begin{list}{}{}
	\scriptsize
  \item[$(1)$] For the Polaris field, the IRAM (small) and
      KOSMA (large) datasets are taken separately. $\dag$:
      based on the \twCO\jtwo\ data, $\ddag$: \twCO\jone\
      data
  \item[$(2)$] the type of PDF ($pos$ or $los$)
  \item[$(3)$] Map sizes (in $\pc\times\pc$) are computed
      assuming a distance of 150~pc
  \item[$(4)$] $\sigma_1$ is the standard deviation
  \item[$(5)$] $\sigma_2$ is derived from a Gaussian fit
  \item[$(6)$] $\sigma_3=\leq/2.35$
  \item[$(7)$] average of the three determinations
  \end{list}
\end{table}

\subsection{Structure functions of the line centroid velocities}

By analogy with studies performed on the velocity field
\citep[\eg][]{she2001}, we computed the structure functions
of the line CV, using the PDF of the centroid velocity
increments \ncpdf, a procedure that allows a check of the
convergence of the structure functions by filtering out
doubtful points in the PDF. The structure functions are
evaluated by a direct integration of the PDF:
\begin{equation}
  S_p(l) = \int_{0}^{\infty} |\dc|^p \absncpdf \,\ud(|\dc|).
  \label{eq:sf}
\end{equation}

Structure functions of velocity are frequently normalized to
the third-order function for two reasons: first because in
incompressible, isotropic, and homogeneous turbulence,
$\zeta(3)=1$ is an exact result; second because oscillations
in $S_p(l)-l$ plots are damped when the $S_p$ are plotted
against $S_3$, a property called extended self-similarity
(ESS) \citep{benzi1993}, so that the range of scales over
which the structure functions are power laws is wider.

Calculations of high-order structure functions are
susceptible to errors since, as $p$ increases, any spurious
large fluctuation largely affects $S_p$. The calculation of
$S_p$ based on Eq.~\ref{eq:sf} allows us to reject bins
populated by too small a number of points, and to study the
influence of irrelevant bins of the PDF (for which
$N<\nmin=10$).  We also applied the procedure described in
\cite{leveque1997} and \cite{padoan2003} for determining the
highest significant order: the peak of the histogram of
$|\dc^p|$ occurs for a value of \dc\ that must be
represented by a significant number of points in the PDF of
\dc. The maximum significant order found is $p=6$.

\begin{figure*}
  \def\wa{0.33\hsize}
  \begin{center}
    \includegraphics[width=\wa,bb=39 116 519 573,angle=0]{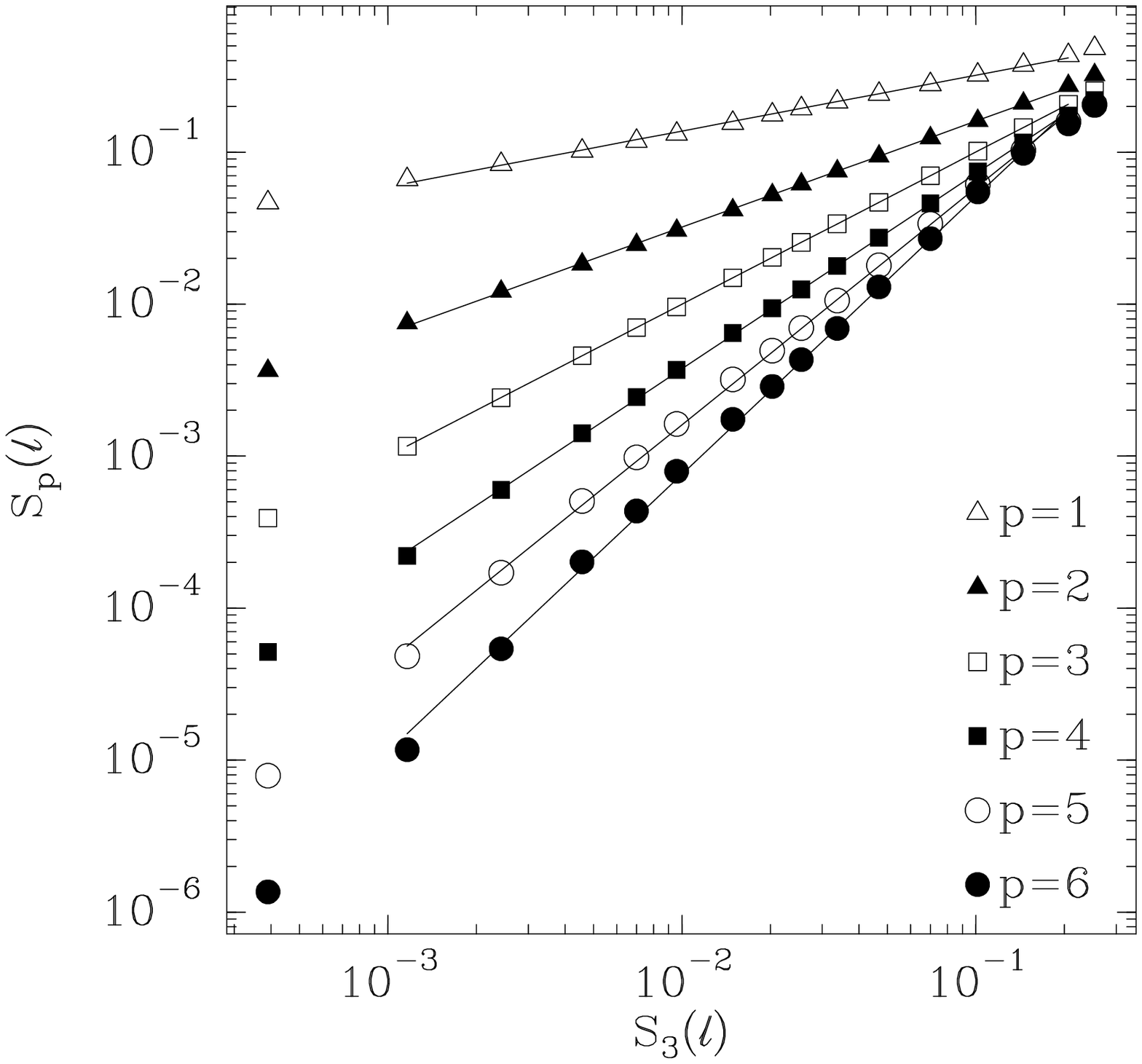}\hfill
    \includegraphics[width=\wa,bb=39 116 519 573,angle=0]{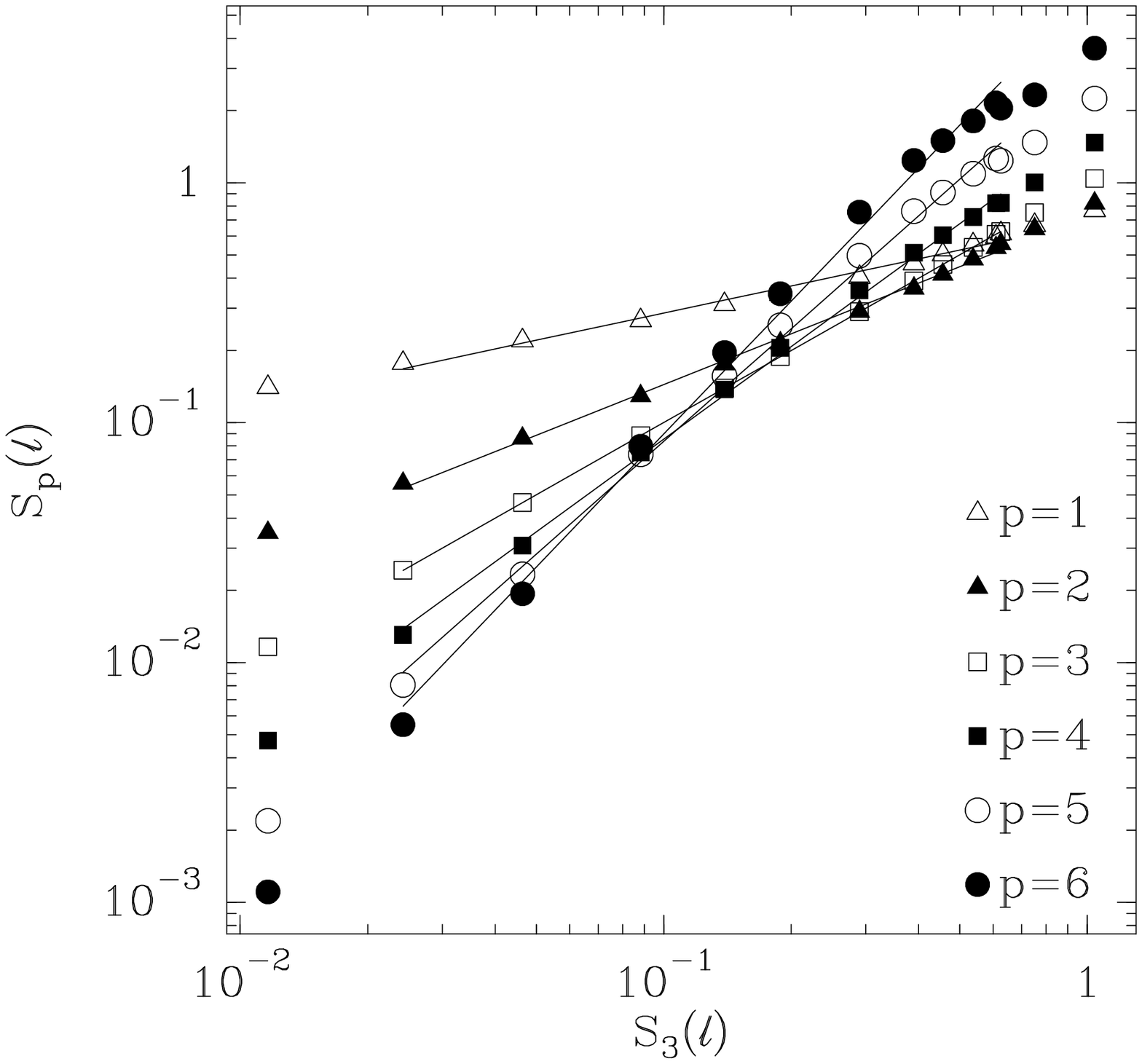}\hfill
    \includegraphics[width=\wa,bb=39 116 519 573,angle=0]{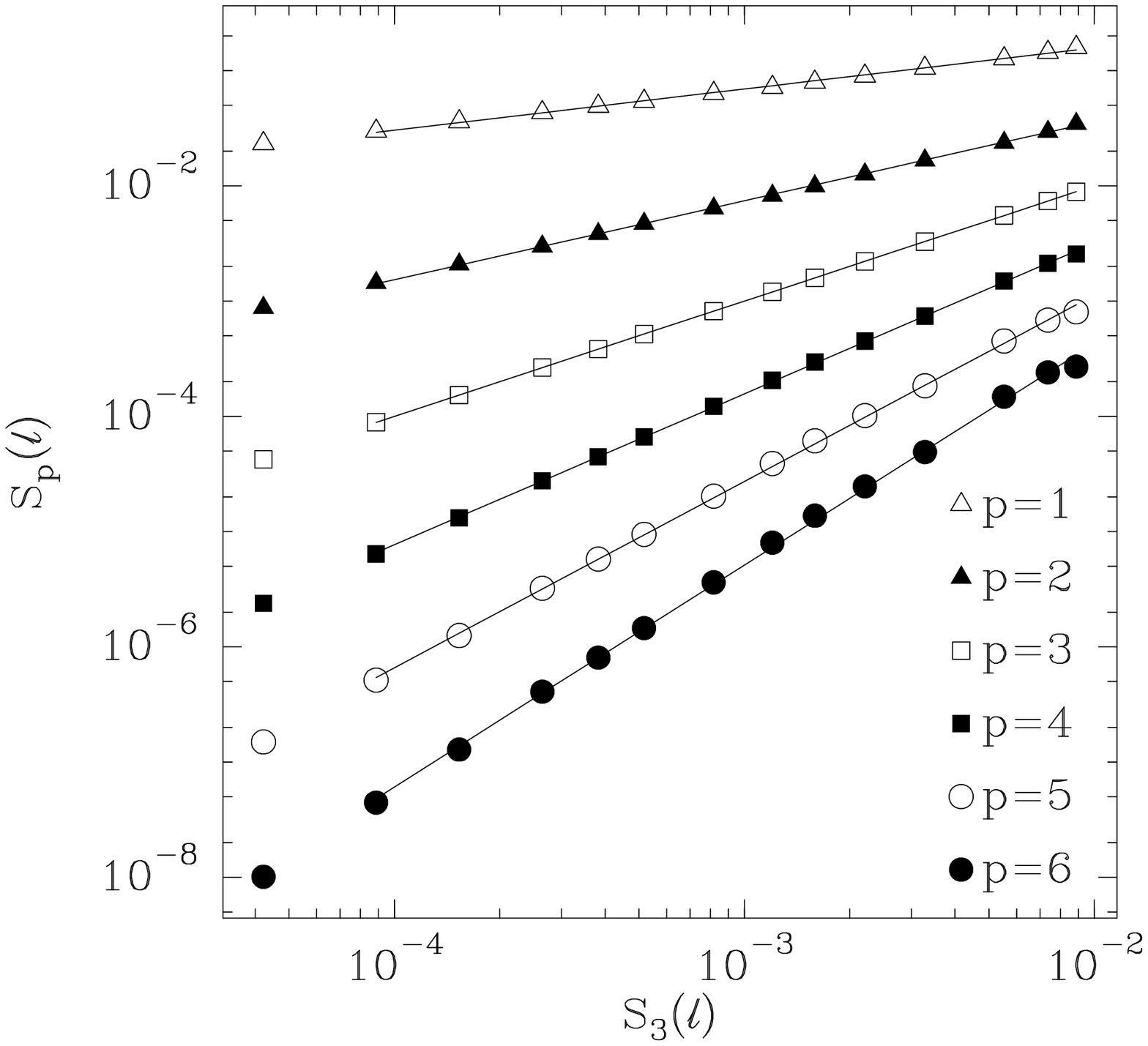}\hfill%
    \caption{Structure functions $S_p(l)$ plotted against
      $S_3(l)$ ($p=1,\ldots,6$) for Polaris (left for the
      IRAM data and middle for the KOSMA data) and Taurus
      (right). A power law is fitted to each order for $l>2$
      pixels, and for $l<60$, $l<30$, and $l<40$ from left
      to right.}
    \label{fig:sfa1}
  \end{center}
\end{figure*}

The scalings of $S_p$ with $S_3$ are shown in
Fig.~\ref{fig:sfa1} for the Polaris and Taurus fields. They
are power laws. Exponents of the structure functions are
calculated by fitting the ESS diagrams for lags wider than 2
pixels and smaller than 30 to 60 (see
Fig.~\ref{fig:sfa1}). Error bars on the exponents ($1-3\%$)
are estimated from weighted averages of the results from two
calculations corresponding to \nmin=1 and 10. The ESS
exponent values are given in Table~\ref{tab:expo}. The
absolute values of the second-order structure function in
the Taurus field are a factor 6 lower than in Polaris,
confirming that the Taurus field contains less kinetic
energy. The non-ESS fits of the CV structure functions lead
to values for $\zeta(3)=1.35$ and 1.60 in Polaris and
Taurus, which differ significantly from the expected value
($\zeta(3)=1$) for the velocity in incompressible,
homogeneous, and isotropic turbulence \citep{kritsuk2007}.

\section{The spatial distribution of the largest line
	  centroid velocity increments}

In the following section, we discuss the spatial structures
of the largest CVI in the different maps. We compare maps of
CVI computed on large and small-scales with large and small
beams. For the sake of simplicity, we call shear the CVI
value divided by the lag over which it is measured,
$\dc/l$. This will be justified at the end of Section~5.

\subsection{Locus of the extrema of CVI in the IRAM fields}

\begin{figure*}[t]
  \begin{center}
    \includegraphics[width=\hsize]{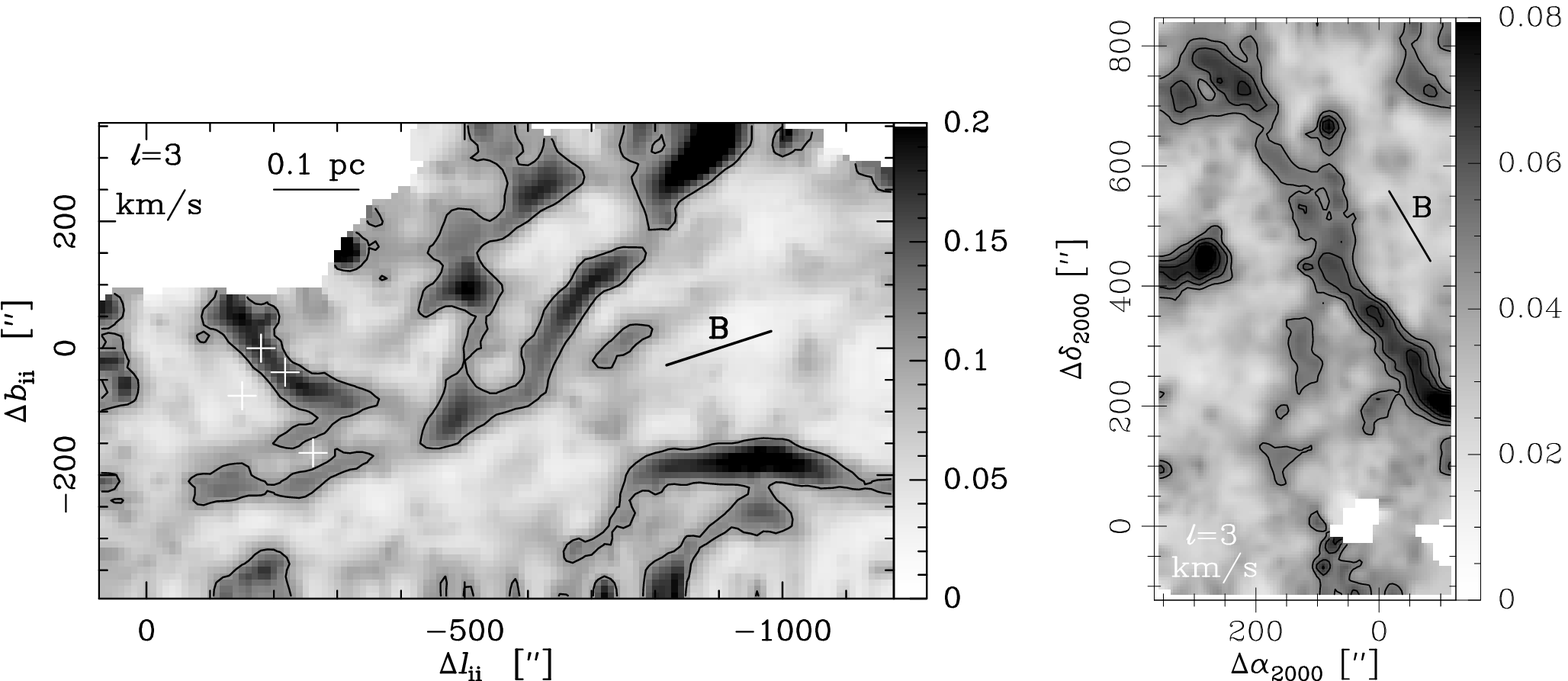}
    \caption{Spatial distribution of the azimuthally
      averaged CVI (\ie\ \avcvi) computed with lags $l=3$
      based on the \twCO\jone\ line. The grey-scale gives
      the \avcvi\ in \kms. The dark regions correspond to
      large values of the CVI associated to the tails of the
      \cpdf\ (Fig.~\ref{fig:pdf-polaris} and
      \ref{fig:pdf-taurus}). The orientations of the
      magnetic fields \citep{heiles2000} are also shown.
      \textit{Left panel}: Polaris field. Contours indicate
      the 0.11 and 0.22~\kms\ levels. The largest CVI
      (0.30~\kms) appear in the NW corner of the field.  The
      4 crosses indicate the positions where the \hcop\
      abundances have been measured (Paper~I). \textit{Right
      panel}: Taurus field. Map center is $\alpha_{2000} =
      \hms{04}{40}{08.84}, \delta_{2000} =
      \dms{24}{12}{48.40}$. Offsets are
      in~\arcsec. Contours: 0.05~\kms\ with 0.01~\kms\
      steps. Note that the largest CVI (0.10~\kms) in that
      field are 3 times smaller than in the Polaris
      field. The blanked areas around $(0,0)$ offsets
      correspond to the positions where the 10~\kms\
      component is present.}
    \label{fig:cvimap-lag3}
  \end{center}
\end{figure*}

Figs.~\ref{fig:cvimap-lag3} and \ref{fig:cvimap-lag18} show
the maps of azimuthally averaged CVI (\avcvi) computed for
two lags ($l=3$ and 18 pixels) in both fields. The grey
scale is the magnitude of the azimuthally averaged CVI at a
given position on the sky.  Because of the averaging
procedure, the exact values of the CVI in those maps are not
trivially related to the values of the non-averaged PDF of
Figs.~\ref{fig:pdf-polaris} and
\ref{fig:pdf-taurus}. However, regions of large CVI in the
maps do correspond to the positions responsible for the
non-Gaussian tails in the PDF of
Figs.~\ref{fig:pdf-polaris}-\ref{fig:pdf-kosma}.

At a small lag ($l=3$), in both fields, the spatial
distribution of the bright regions with large CVI delineates
elongated structures. When the lag is larger (18 pixels),
the contrast of the structures above the background values
fades away. Yet, in the Polaris field, the structure around
$(-1000\arcsec,-200\arcsec)$ is still visible with $l=18$,
and for the two lags of 3 and 18 pixels, the largest CVI is
located in the northwestern corner of the map.

\begin{table}[t]
  \begin{center}
    \caption{Exponents $\zz_p=\zeta(p)/\zeta(3)$ of the ESS
      structure functions of the CV for the Polaris and
      Taurus fields (see Fig.~\ref{fig:sfa2}).}
    \begin{tabular}{l c c c c c c}\hline
      \vtab{0}{3}%
      & $\zz_1$ &$\zz_2$ &$\zz_3$ &$\zz_4$ & $\zz_5$ & $\zz_6$ \\\hline
      Polaris         & 0.37 & 0.70 & 1.00 & 1.27 & 1.53 & 1.77 \\
      Polaris$^{(a)}$ & 0.38 & 0.71 & 1.00 & 1.28 & 1.54 & 1.80\\
      Taurus          & 0.36 & 0.69 & 1.00 & 1.30 & 1.60 & 1.89 \\
      SL94$^{(b)}$    & 0.36 & 0.70 & 1.00 & 1.28 & 1.54 & 1.78 \\
      B02$^{(b)}$     & 0.42 & 0.74 & 1.00 & 1.21 & 1.40 & 1.56 \\
      \hline
    \end{tabular}
	\begin{list}{}{}
	  \scriptsize
	  \item[$(a)$] From the large scale \twCO\jtwo\ data
      from \cite{bensch2001}.
	  \item [$(b)$] SL94 and B02 are the \cite{she1994} and
      \cite{boldyrev2002} scalings of the velocity structure
      functions (see Section~5.2).
	\end{list}
    \label{tab:expo}
  \end{center}
\end{table}
These maps show that the positions of the largest CVI, in
Polaris and Taurus, are not randomly distributed but are
connected and form elongated structures. In the Taurus
field, the direction of the most prominent CVI structure is
parallel to the projected orientation of magnetic fields
measured in the NE corner of the field \citep{heiles2000}
(see also Fig.~\ref{fig:polaris-iris}). In Polaris, the
scatter of their orientations relative to the magnetic field
is larger (Paper~II). In both Taurus and Polaris, their
characteristic half-maximum width, measured on transverse
cuts, is resolved (30\arcsec\ after deconvolution from the
lag, or $\dfil\approx0.02$~pc), and their aspect ratio is
often greater than 5. The surface fractions covered by the
regions where the \avcvi\ are larger than 3$\sigaver$ (where
\sigaver\ is the dispersion of the distribution \avcvi), are
10\% and 28 \% in Polaris and Taurus, respectively.  Last,
the non-averaged CVI in these structures (see Section 3.1)
are 5 and 4 times larger than the dispersion \sigdc\ (see
Figs.~\ref{fig:pdf-polaris} and \ref{fig:pdf-taurus}) in
Polaris and Taurus, respectively. For $l=3$, the
corresponding shears are $5\times0.11~\kms / 0.02~\pc =
30~$\kmspc\ in Polaris and $4\times0.05~\kms /
0.02~\pc=10~$\kmspc\ in Taurus. The most turbulent field on
the parsec-scale (Polaris) is therefore that where the
largest small-scale shears are measured.

\subsection{Comparison of the extrema of CVI with the CO emission}

We stress here that our work is based on the statistics of
the extrema of CVI (called E-CVI in what follows), unlike
what is done in most analyses \citep[\eg][]{esquivel2005},
where the full distribution of CVI is considered.  We show
below which specific features of the CO line emission are
associated with the extrema of CVI.

\begin{figure}
  \begin{center}
	\includegraphics[width=0.47\hsize,angle=0]{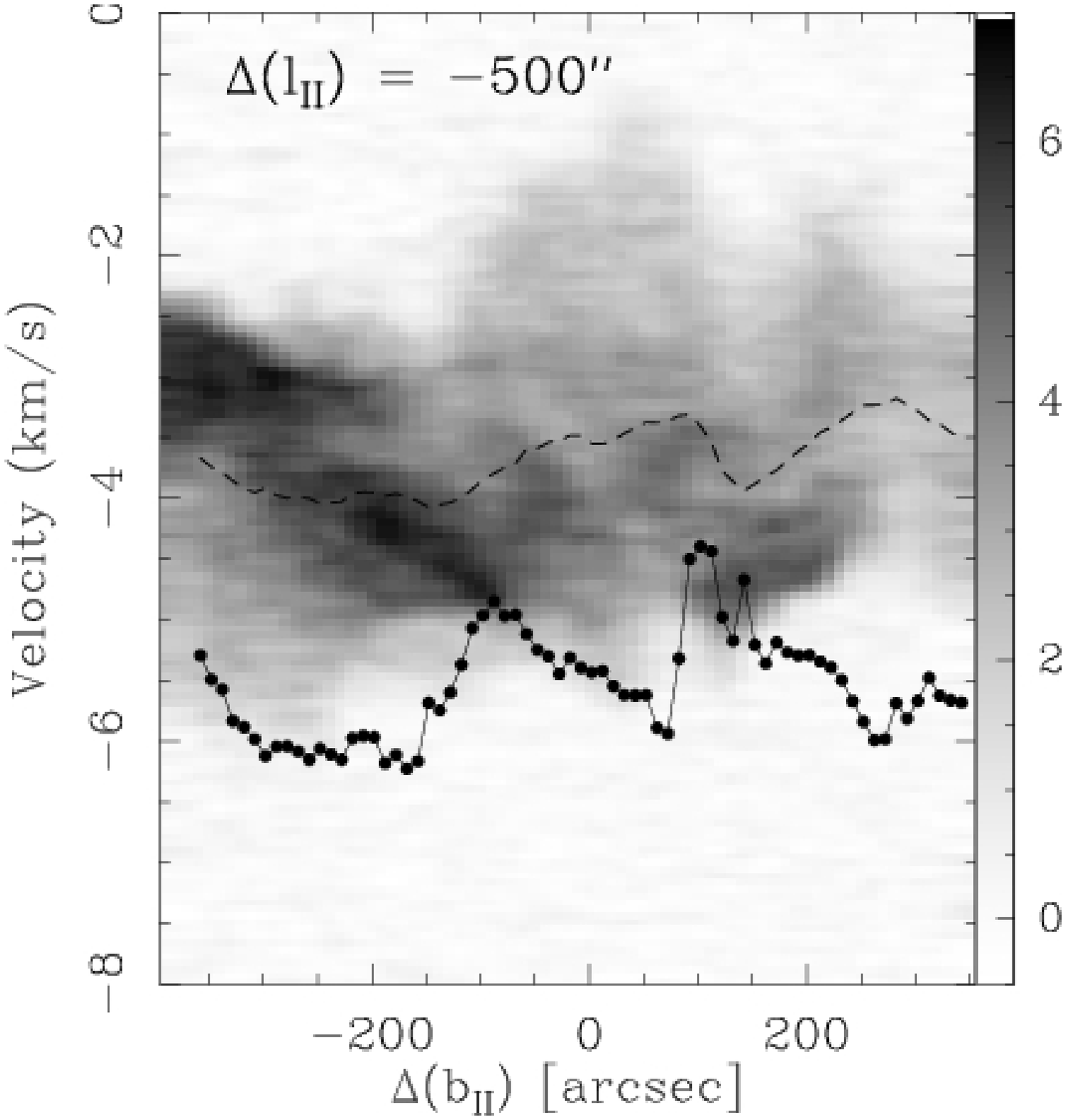}\hfill%
	\includegraphics[width=0.47\hsize,angle=0]{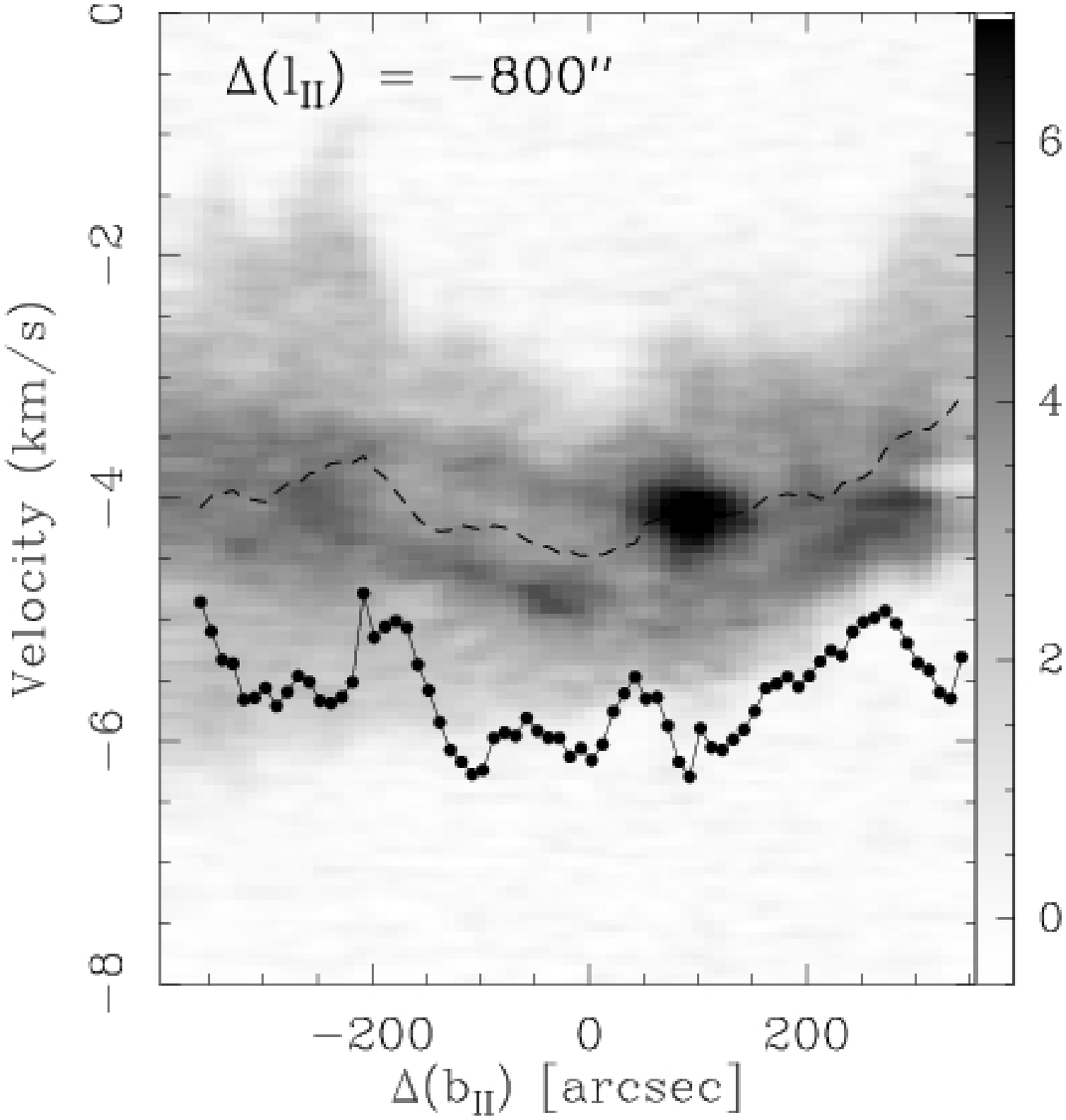}
	\caption{Position-velocity cuts at constant \lii\
	  offsets ($\Delta\lii=-500$\arcsec, left and
	  $\Delta\lii=-800$\arcsec, right). \emph{Grey-scale}:
	  main-beam intensity in K. \emph{Dashed curve}:
	  centroid velocity. \emph{Full curve}: azimuthally
	  averaged CVI for a lag $l=3$ pixels (with an
	  additional offset of -6.5~\kms\ for clarity).}
	\label{fig:lvcut}
  \end{center}
\end{figure}

Fig.~\ref{fig:lvcut} displays two \twCO\jone\ space-velocity
cuts made across the Polaris map at longitude offsets -500
and -800\arcsec.  The run of the averaged CV and CVI along
these cuts is shown to illustrate that the largest CVI are
mostly due to very localized broad CO linewings. As
expected, however, some of the large variations in the
centroid velocities due to these broad wings are reduced by
opposite variations due to fluctuations in the line-core
emission.
 
This is seen better in Fig.~\ref{fig:12co10-cvi} where the
extrema of centroid velocity increments in the Polaris field
are overplotted on the \twCO\ wing emission and the \thCO\
integrated emission. The \twCO\ wing emission (top panel) is
optically thin and associated with warm and tenuous gas
emission \citepalias{hilyblant2007ii}. The \thCO-integrated
emission (bottom panel) is used here as a proxy for the
molecular gas column density. While the largest CVI are not
spatially correlated with the \thCO-integrated emission,
they closely follow the boundaries of the optically-thin
\twCO\ emission in the broad \twCO\ linewings.

Given the high latitude of the Polaris cloud, the structures
responsible for the broad \twCO\ linewings most likely
belong to the Polaris Flare
(Fig.~\ref{fig:polaris-iris}). The non-Gaussian tails of the
PDF at small lags are thus associated to local structures on
the scale of 30~pc. Their distance, and therefore their
size, is known to within 20\%, while the lags of the PDF
decrease by an order of magnitude (from 25 to 3). The $l=3$
pixel PDF is thus sensitive to true small-scale structures
created by the turbulence in the Polaris Flare.  This
comparison shows that the regions of largest shear are not
associated with the bulk of the condensed matter in the
field, but are instead correlated with warmer and more
diluted gas.  We therefore infer that, unlike centroid
velocities increments in general, the E-CVI we analyze are
not due to density fluctuations or radiative transfer
effects in optically thick gas.

We now address the issue of the chance coincidence of
unrelated pieces of gas on the line of sight.  In the two
translucent fields, CO is not expected to trace the full
molecular content of the clouds, essentially as a result of
photodissociation processes.  This has possibly been
observed by \cite{sakamoto2003} who show discontinuous CO
emission in low extinction regions in the Taurus
complex. These spots of CO emission are however embedded in
the underlying turbulent molecular gas, undetected because
mostly made of molecular hydrogen, and presumably
continuous. The velocity field deduced from the CO emission
lines thus carries the statistical properties of that
turbulent molecular gas.
 
Nonetheless, projection effects are inevitable, and a key
parameter is the ratio $l/L$ of the lag $l$ over which CVI
are measured to the unknown depth of the cloud along the
line-of-sight $L$. In their work on 512$^3$ numerical
simulations of mildly compressible turbulence,
\cite{lis1996} have computed PDF of CVI for a lag of 3
pixels, for which this ratio is 3/512=0.006. They show that
the E-CVI trace extrema of $\aver{ (\rot{v})_y }_{\rm los}^2
+ \aver{(\rot{v})_z}_{\rm los}^2$. Since this quantity is a
$los$ integration of a signed quantity (the two projections
of the vorticity in the plane of the sky), its extrema are
due to a few exceptional values present on the line of
sight. For this reason \cite{lis1996} say that the E-CVI
trace the projection of large velocity-shears (or vorticity)
in turbulence.  In Polaris and Taurus, $L$ is not known but
we conservatively adopt a value in the range $1-30$~pc. For
the smallest lag ($l=3$ pixels) the ratio is in the range
$l/L=0.001-0.02$. Our observational study thus falls into
the regime tested by \cite{lis1996}, and the filaments
associated with the E-CVI thus trace the projection of
regions of extreme velocity-shear somewhere on the line of
sight.  The observed value of the shear, though, is of
course an upper limit.  We are therefore confident that
E-CVI trace genuine extrema of line-of-sight velocity
fluctuations.

\subsection{Parsec-scale coherence of the regions of largest CVI: 
IRAM and KOSMA data}

We compare here the properties of the E-CVI (values and
spatial distribution) from the KOSMA and IRAM data sets of
the Polaris field. The comparison is not, however,
straightforward for two reasons. First, the value of the
centroid velocity is affected by the beam size, and second,
the computation of the CVI filters out any structures that
are much larger than the lag.

\begin{figure}[t]
  \begin{center}
    \includegraphics[width=0.9\hsize]{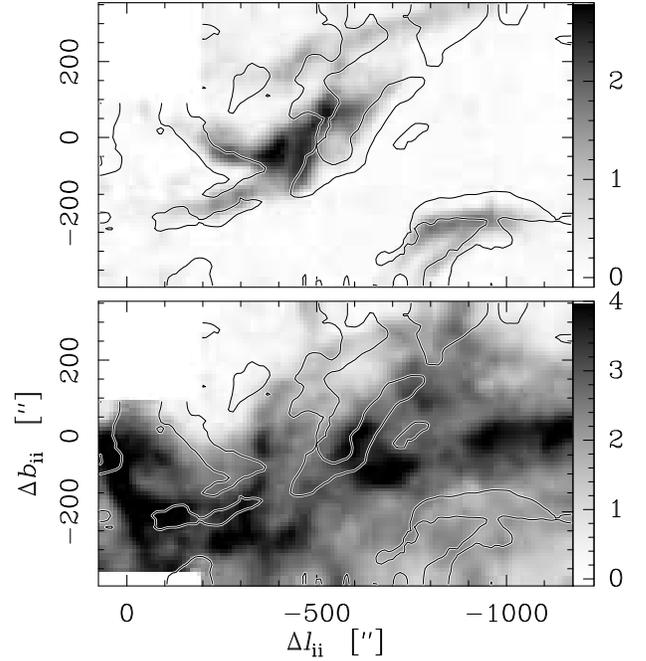}
    \caption{Contours of the CVI (1~\kms) computed in the
      Polaris field for $l=3$ pixels (see
      Fig.~\ref{fig:cvimap-lag3}) overplotted on the
      \twCO\jone\ emission integrated in the velocity range
      $[-2,5:0]$~\kms\ \ie\ optically thin \twCO\ emission
      (top) and the \thCO\ integrated intensity tracing the
      bulk of the dense gas (see Paper~II) (bottom).}
    \label{fig:12co10-cvi}
  \end{center}
\end{figure}

\begin{figure}
  \begin{center}
    \includegraphics[width=\hsize]{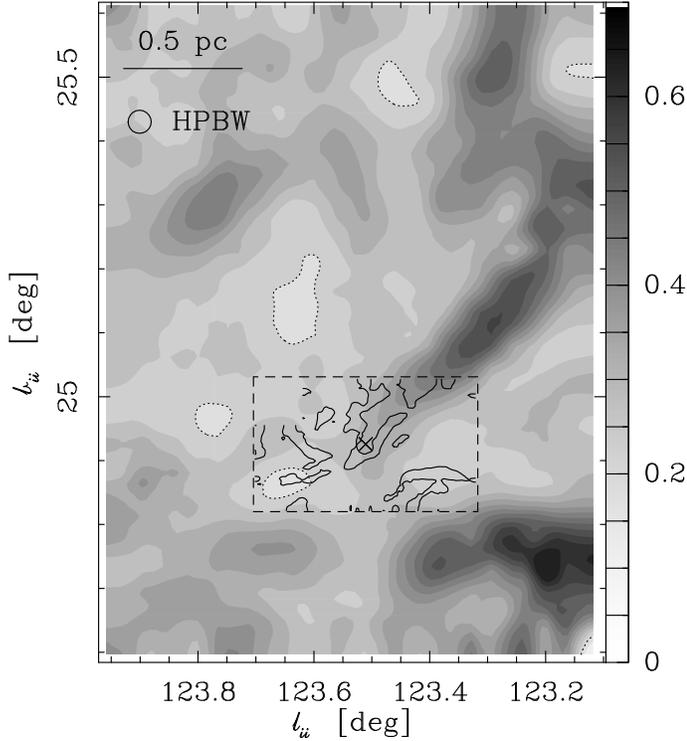}
    \caption{Spatial distribution of CVI (\kms) computed on
      large scales using the data of \cite{bensch2001}
      (grey-scale with levels indicated on the right scale,
      dotted contours indicate the 0.1~\kms\
      level). Full-line contours indicate the CVI computed
      on small scales in the IRAM-30m data cube (see
      Fig.~\ref{fig:12co10-cvi}). Note the spatial
      coincidence of the IRAM structures at the tip of the
      prominent KOSMA structure at (123.4\degr, 25\degr),
      where the CVI are 0.30~\kms.}
    \label{fig:cvikosma}
  \end{center}
\end{figure}

Fig.~\ref{fig:cvikosma} displays the spatial distribution of
the CVI computed on large scales with the KOSMA data, for a
lag $l=3$ pixels (180\arcsec). Regions of large increments
are spatially resolved filaments: cuts across the structures
provide an average thickness of 200\arcsec\ deconvolved from
the lag, or 0.18~pc. These structures are about 7 times
thicker than those found in the IRAM field. The prominent
KOSMA structure around (123.29\degr, 25.11\degr) smoothly
connects with the northwestern IRAM structure (contours from
Fig.~\ref{fig:12co10-cvi}). This is seen more clearly in
Fig.~\ref{fig:cvicut} where the values of the non--averaged
CVI along this structure are displayed.
Fig.~\ref{fig:cvicut} illustrates three points: \textit{i)}
the CVI from the KOSMA and IRAM datasets decrease
monotonously from north to south along this structure,
\textit{ii)} the IRAM CVI measured over $l=180$\arcsec\ are
all larger than those measured with KOSMA over the same
physical lag, and \textit{iii)} the CVI measured with the
KOSMA telescope over a lag of 3 pixels (180\arcsec) are
similar to those measured at the same positions with the
IRAM telescope with a lag 6 times smaller (30\arcsec) and .

The latter property is unexpected: it suggests that the CVI
are similar in this region whether they are measured with a
small beam and a small lag (IRAM) or a large beam and a
large physical lag. If the KOSMA structures were only due to
beam-dilution of the IRAM ones, the KOSMA CVI for
$l=180$\arcsec\ would be smaller than the IRAM ones for
$l=30$\arcsec. In other words, the KOSMA CVI structures are
real and are sub-structured: the same velocity variations
($<0.5$~\kms\ for positive offsets) are measured on small
(IRAM) and large (KOSMA) scales.

The largest velocity-shears at the KOSMA resolution are
$5\times0.30 / 0.18 \approx 8.3$~\kmspc. The surface
fraction covered by the large increment structures where
$\avcvi>3\sigaver$ are close, 10 and 16\% for the IRAM and
KOSMA data, respectively. These two fields, observed with
different telescopes and resolutions, thus show similar
statistical properties and demonstrate the coherence at the
parsec-scale of the structures of largest centroid velocity
increments. This is discussed in the next section.


\section{The two facets of intermittency: statistical and structural}

In Section~3, the two-point statistics of the line centroid
velocities were found to display marked non-Gaussian
behaviors. In Section~4, the emission responsible for these
non-Gaussian statistics is resolved into coherent
structures. We here compare these statistical and structural
characteristics with theoretical predictions and recent
numerical results regarding the intermittency of turbulence.

\subsection{Self-similarity of the centroid velocity increments  
in the Polaris field}

The IRAM and KOSMA PDF from Figs.~\ref{fig:pdf-polaris} and
\ref{fig:pdf-kosma} bear an apparent contradiction: the PDF
built with the IRAM data with a lag of 150\arcsec\ (15
pixels) is nearly Gaussian, while the KOSMA PDF with a
similar physical lag of 180\arcsec\ (3 pixels) is not:
non-Gaussian tails in the KOSMA PDF originate from the
largest CVI structures like the most prominent one discussed
in Section~4.3.  In Fig.~\ref{fig:cvicut}, we show that the
bulk of the IRAM CVI for $l=180$\arcsec\ (dark dots) are
below $3\sigdc=0.9$\kms.  It is the particular location of
the IRAM field with respect to the CVI maxima seen in the
KOSMA field that prevents the detection of a number of CVI
in excess of 3\sigdc\ large enough to depart from Gaussian
statistics. Would the IRAM field be centered closer to the
CVI maximum in the KOSMA data (around 123.29\degr,
25.11\degr), a larger number of CVI in excess of 3\sigdc\
might have been measured.

However, in both fields, the non-Gaussian tails of the PDF
of CVI grow as the lag decreases. This behavior is routinely
observed in laboratory and numerical experiments, where it
is interpreted as a signature of the intermittency of the
velocity field. In such experiments, effects of the
non-Gaussian statistics are visible with either the
transverse or the longitudinal velocity increments
\citep{frisch1995, mininni2006a}. The present analysis shows
various degrees of non-Gaussianity: the statistics in Taurus
are nearly Gaussian (flatness close to 3), while both
Polaris data sets show clear departure from Gaussianity. If,
following PF03, we attribute the non-Gaussian statistics of
the CVI to the intermittency of the turbulence in the two
sampled molecular clouds (see below), the new result, here,
is that intermittency is as pronounced at a lag
$l=180$\arcsec\ in the large field (KOSMA PDF) as it is at
lag $l=30$\arcsec\ in the IRAM field. In both datasets, the
non-Gaussian tails extend to $5.5-6\sigdc$. This suggests
that intermittency is not only a small-scale phenomenon but
that it is present and has the same statistical properties
on a scale six times larger.

As mentioned in the introduction, similar conclusions have
been reached by MJ04, who find that the intense structures of
vorticity and rate of strain in hydrodynamical turbulence
form clusters of inertial range extent, implying a
large-scale organization of the small-scale intermittent
structures.

\subsection{The intensity of small-scale intermittency 
versus large-scale shear}

We now compare the scaling of the $p^{th}$-order structure
functions of CV with $p$ in the framework of the SL94
model. Structure functions are increasingly sensitive to the
tails of the CVI PDF (E-CVI) as the order $p$
increases. Since we have shown (Sect~4.2) that the E-CVI
stem from velocity fluctuations, it is interesting to
compare the scaling of high order structure functions of the
CV to theoretical predictions based on the velocity field.

The SL94 model has three parameters (see Appendix~A.3). One
of the three parameters, $0\le\beta\le1$, describes the
level of intermittency: $\beta \rightarrow 1$ corresponds to
the non-intermittent cascade with $\zz_p = p/3$. The two
other parameters \citep{boldyrev2002} describe the scalings
of velocity in the cascade and the dimension $D$ of the most
intermittent structures.  In the SL94 model, the scaling of
the velocity is $v_l \sim l^{1/3}$. It assumes that the most
intermittent structures are filaments ($D=1$) and that the
level of intermittency is $\beta=2/3$. The associated ESS
structure function exponents $\zz_p = \zeta(p)/\zeta(3)$ are
then predicted to be $\zzsl = p/9 + 2[
1-(2/3)^{p/3}]$. According to this class of models, as the
level of intermittency increases, the ESS exponents become
smaller than $p/3$ for $p>3$ and the departure from the K41
scaling increases with $p$. Following the SL94 approach,
further theoretical models were developed for compressible
and magnetized turbulence \citep{politano1995a,muller2000}
and tested against numerical simulations.
\cite{boldyrev2002} propose a similar scaling to describe
compressible MHD turbulence assuming sheet-like intermittent
structures ($D=2$) and a more intermittent cascade
($\beta=1/3$) but do not allow dissipation of large-scale
modes in shocks. They find $\zzmhd = p/9+1-(1/3)^{p/3}$
(hereafter called the B02 scaling), in excellent agreement
with numerical simulations of super-Alfv\'enic MHD
turbulence. However, this scaling has never been tested
against observations of the turbulent velocity field of
molecular clouds.

\begin{figure*}[t]
  \def\wa{0.3\hsize}
  \begin{center}
	\includegraphics[width=\wa,bb=39 116 519 573,angle=0]{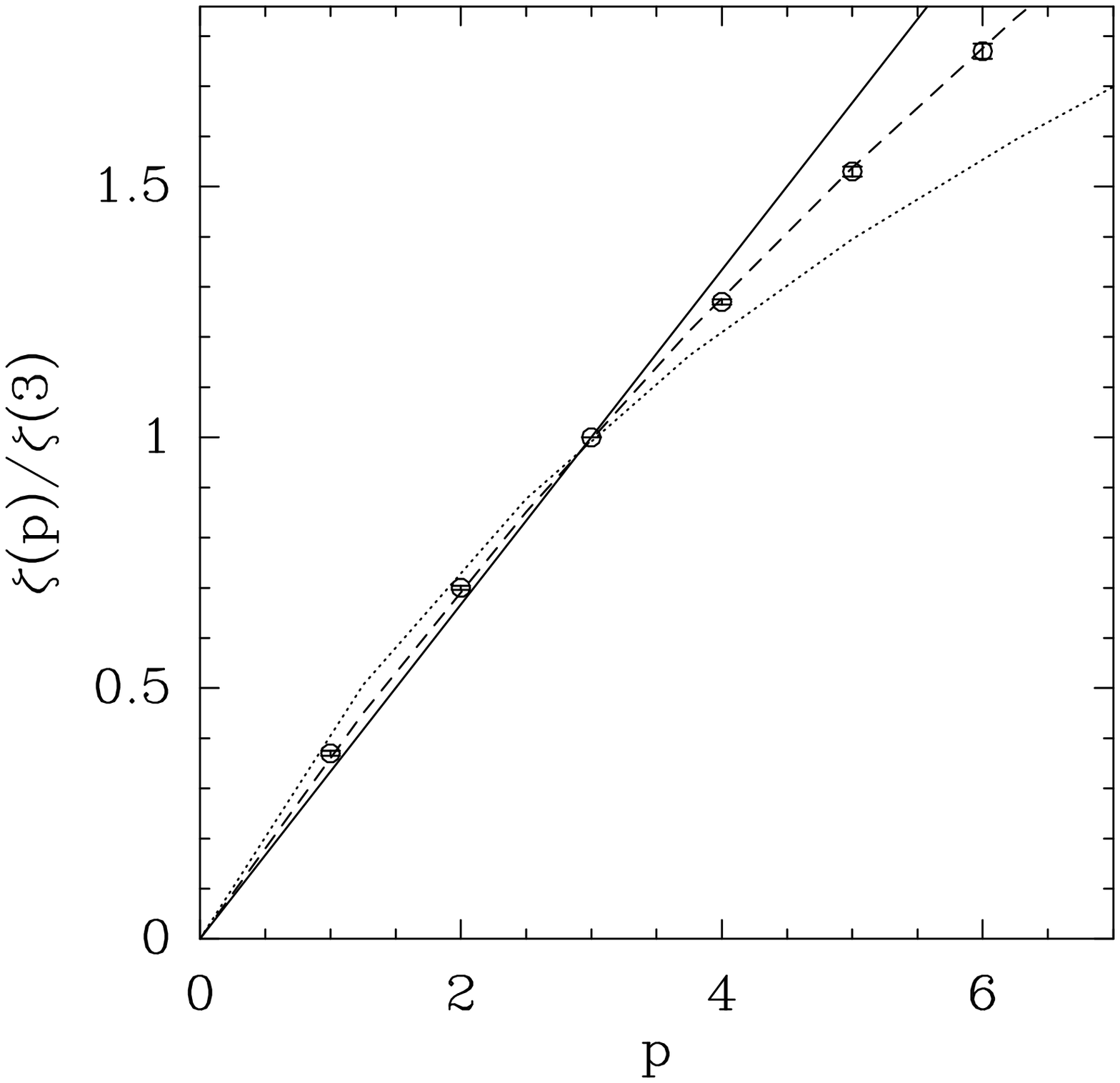}\hfill%
	\includegraphics[width=\wa,bb=39 116 519 573,angle=0]{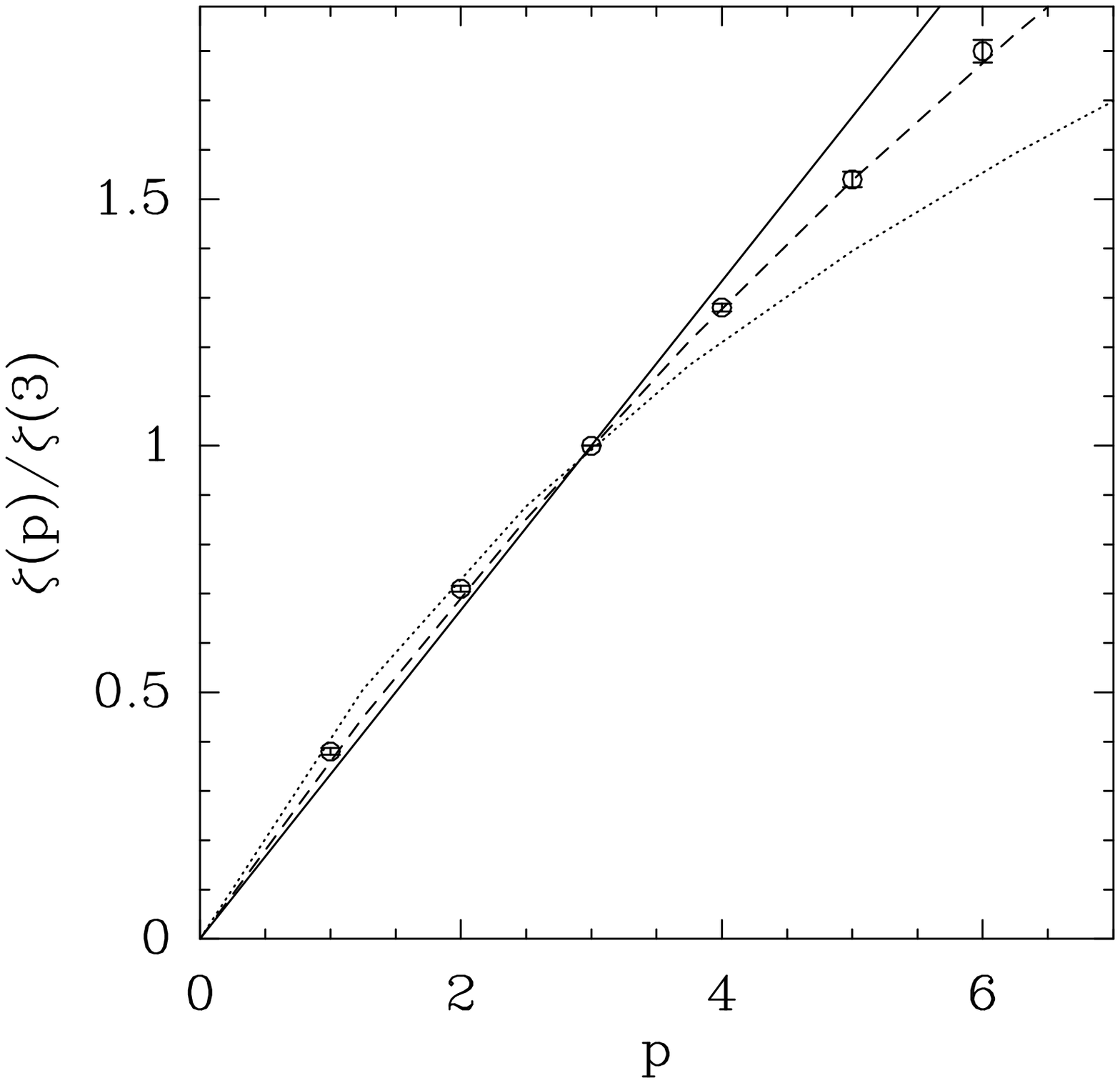}\hfill%
	\includegraphics[width=\wa,bb=39 116 519 573,angle=0]{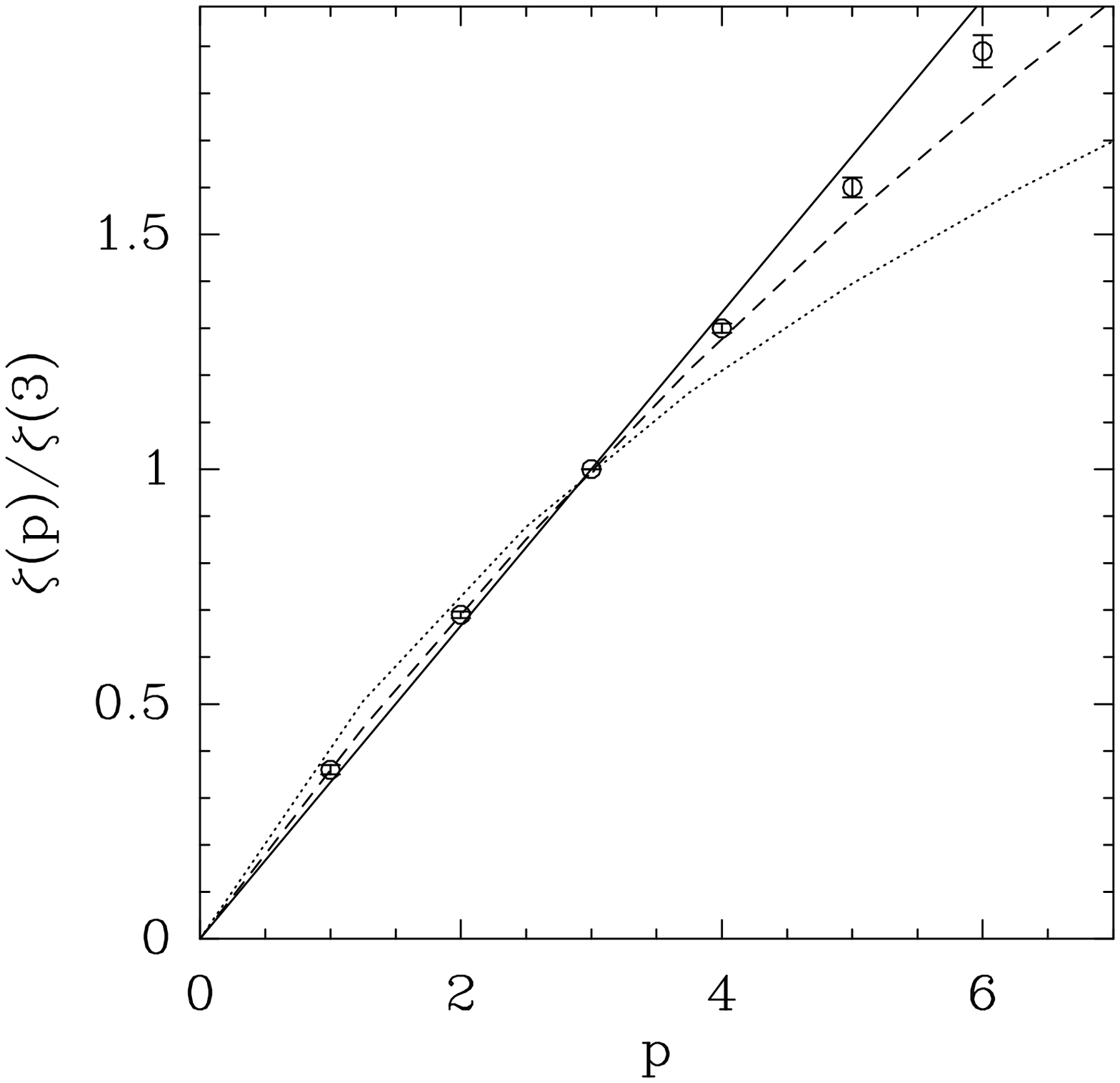}
	\caption{Values of the exponents of the structure
	  functions (see Fig.~\ref{fig:sfa1}) with error bars
	  (see Table~\ref{tab:expo}). For comparison, the K41
	  (full), SL94 (dashed), and B02 (dotted) scalings are
	  indicated. From left to right: Polaris (IRAM), Polaris
	  (KOSMA), Taurus.}
	\label{fig:sfa2}
  \end{center}
\end{figure*}

In Fig.~\ref{fig:sfa2}, we compare the scalings of the CV
structure functions exponents (see Table~\ref{tab:expo})
with the predictions of the SL94 and B02 models for the
velocity field. In the Polaris field, for either data set,
the measured exponents follow the SL94 scaling closely but
differ significantly from that of B02.  This apparent
agreement with the SL94 scaling is unexpected since this
model describes incompressible and unmagnetized turbulence,
everything interstellar turbulence is not. However, the
effect of line-of-sight averaging in the CV structure
functions is not known. Further interpretation of the
underlying physics requires confrontation with CV structure
functions based on numerical simulations. The exponents in
the Taurus field do not follow any of the three scalings,
and are halfway between the non-intermittent K41 scaling
($\zz(p)=p/3$) and the SL94 scaling.  The Taurus field is
thus less intermittent than the Polaris one, a result
consistent with the flatness measure of the non-Gaussianity
of the \cpdf.  This result is also in line with the recent
numerical findings of \cite{mininni2006a}, who show that the
characteristics of the large-scale flow play an important
role in the development of small-scale intermittency and
determine its statistical properties.

The use of CV structure functions to determine the three
parameters of the SL94 class of models not only requires a
large number of data points but also ``calibration'' of the
weighting performed by CV upon the velocity field, using
numerical simulations of compressible turbulence. The values
that we have determined from the exponents $\zz(p)$ may be
useful, though, and we give them in the Appendix.

In summary, the line CV exhibit the statistical and
structural properties characterizing the intermittency of
the velocity field in theoretical models and numerical
simulations of turbulence: 1) the non-Gaussian statistics of
the CVI at small lag, 2) the self-similarity of structures
of largest CVI and the existence of inertial-range
intermittency, 3) the anomalous scaling of their high-order
structure functions similar to wthat is found for the
structure functions of velocity. Last, we find that the more
intermittent field on small scales has the larger dispersion
of non-thermal velocity on large scales.

The above properties are borne by the non-Gaussian tail of
the CVI PDF, which we have shown to be associated with pure
velocity fluctuations. They support our proposition that
statistics of the \twCO\ line centroid velocity and, more
specifically, their E-CVI may be used to disclose the
statistical and structural properties of intermittency in
the underlying velocity field. In what follows, we therefore
ascribe the E-CVI to the intermittent structures of intense
shears.

\section{Discussion}

\subsection{Influence of gravity}

To test the role of gravity in the generation of
non-Gaussian statistics of the velocity, \cite{klessen2000}
built the two-point statistics of the velocity field, in SPH
numerical simulations of turbulence, both with and without
self-gravity .  The author compared the numerical PDF of CVI
with observed PDF and concluded that the inclusion of
self-gravity leads to better agreement with the observed PDF
in molecular clouds. It was further argued in PF03 that this
result was indeed expected since the observed regions used
in the comparison are forming stars, hence the importance of
self-gravity. The situation is drastically changed in the
two translucent fields we have studied: they do not form
stars, are far from any such regions, and both are
non--self-gravitating on the parsec-scale of our
observations. Indeed, we have shown that the field with the
more prominent non-Gaussian tails, namely Polaris, is
located at high latitude and is embedded in a larger
structure (the Flare) far from virial balance. This strongly
supports that, in the type of fields we analyze, gravity is
not at the origin of the PDF tails.

Nonetheless, gravity is the ultimate source of gas motions
in the universe, from galaxy clusters to GMC and collapsing
cores, so it cannot be ignored. If the cloud mass were
distributed in tiny cloudlets of very high density that
would rarely collide, then gravity might play a significant
role in the gas velocity statistics.  Here, we assume that
the fluid approximation is valid, and because the Reynolds
number is so large, the gas motions are turbulent, by
definition.

\subsection{The intermittency of turbulence dissipation}

\begin{figure}
  \begin{center}
	\includegraphics[width=0.8\hsize]{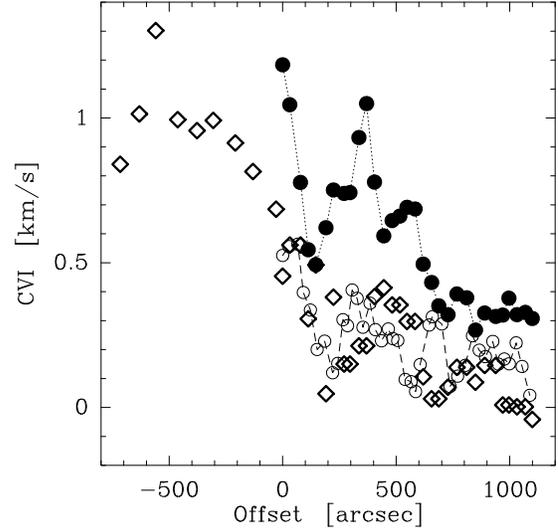}
	\caption{Values of the CVI in the Polaris field, along
	  the NW-SE CVI structure from (123.2, 25.2~\degr) to
	  (123.6, 24.9~\degr) seen in
	  Fig~\ref{fig:cvikosma}. CVI are from KOSMA data for
	  $l=3$ (diamonds), IRAM $l=3$ (open circles), and IRAM
	  $l=18$ (filled circles). Offsets increase from NW to
	  SE, with the zero position corresponding to the NW
	  corner of the IRAM field (see
	  Fig.~\ref{fig:cvimap-lag19-pol}).}
	\label{fig:cvicut}
  \end{center}
\end{figure}

In numerical experiments such as those of MJ04 \citep[see
also][]{sreenivasan1999}, maxima of energy dissipation are
found, at small-scale, in the vicinity of the vorticity
filaments.  A large fraction of the dissipation of
turbulence may be concentrated in the regions of largest
CVI, the small-scale intermittent structures.  We illustrate
this point with estimates of energy transfer based on our
observations in the Polaris field.

The transfer rate in the cascade on scale $l$ is $\dissip{l}
= 1/2 \rho \vcar{l}^3/l$ with $\vcar{l}$ the characteristic
turbulent velocity fluctuations on that scale. This assumes
that the time to transfer energy from scale $l$ to smaller
scales writes as $\tau_l = l/\vcar{l}$. At the parsec-scale
of the Polaris cloud
\begin{equation}
  \dissip{L} = 5.5 \tdix{-25} / L_{\rm pc}^2 \qquad \emiss
  \label{eq:dissipL}
\end{equation}
where the average density $\rho=\mu N_H /L$ and the velocity
dispersion $\vcar{L}=1.5$~\kms\ are those derived in
Paper~II. There, $L_{\rm pc}$ is the unknown depth along the
line of sight, expressed in pc.  The CO cooling rate
averaged over the whole field on the parsec-scale is
(Paper~II)
\begin{equation}
  \overline \Lambda_{\rm CO}= \dix{-24} /L_{\rm pc} \qquad \emiss.
  \label{eq:lco}
\end{equation}
These two rates are thus close, because the depth along the
line of sight is not much greater than 1~pc (see
Section~3.1), and may lead to the conclusion that CO is able
to radiate the turbulent energy away, confirming the results
of \cite{shore2006}.  Actually, this would be true if the
cascade were filling space uniformly or, in other words, for
a non-intermittent cascade. From the previous section, this
assumption is certainly not valid. Indeed, the cumulative
distribution of $\dc^2$ for $l=3$ (see
Fig.\ref{fig:pdf-dissip}) shows that the points with CVI
larger than 3\sigdc\ represent only 2.5\% of the total,
while they contribute to 25\% of the total of $\dc^2$. The
velocity field has two contributions (solenoidal and
dilatational) to the energy dissipation rate $\epsilon =
-\frac{1}{\re} \, ({|\rot{v}|^2+\frac{4}{3}|\div{v}|^2})$
\citep{kritsuk2007}. As shown by \cite{lis1996} and
\cite{pety2000} based on numerical simulations, the E-CVI
serve as a proxy for large vorticity regions. Thus, assuming
that the energy dissipation on scale $l$ is proportional to
$\dc^2$, the cumulative distribution of
Fig.~\ref{fig:pdf-dissip} suggests that the local
dissipation rate at $l=3$ (or $\sim 0.02$ pc) in the E-CVI
regions, is already 10 times larger than the average rate
over the field, $\dissip{\rm E-CVI}>10 \dissip{L}$.  Note
that these numbers are about the same for the large-scale
Polaris field, while in the less intermittent Taurus field,
these E-CVI represent only 1\% of the total, still
contributing to 5\% of the dissipation.

On the actual dissipation scale, presumably smaller than
0.02~pc, the local dissipation rate is still higher by an
unknown factor. Since the turbulent dissipation is not
space-filling, it induces high local heating rates. This
suggests that other cooling agents, \eg\ the pure rotational
lines of \hh\ or the fine structure line of \cp, may be
dominating the cooling in these regions \citep[Paper~I
  and][]{falgarone2007}. Observations are still lacking that
would allow a comparison of the turbulent transfer rate with
the CO cooling rate on scales smaller than 0.02~pc.

In spite of the self-similarity of the intermittent
structures discussed in Section~5, the bulk of the
dissipation is likely to take place in the smallest
structures. The largest CVI are proportional to \sigdc\ and
thus to $l^{1/2}$. The corresponding shears therefore scale
as $l^{-1/2}$ providing an observed scaling of the
dissipation rate with lengthscale $l$, $\epsilon_l \propto
l^{-1}$. Now, we use the finding of MJ04, who show that the
tails of the probability distribution functions of the
volume of individual dissipative structures (either intense
vorticity or strain-rate) decrease approximately as $p(V)
\sim V^{-2}$. Whether these structures are cylinders ($V
\propto l^2$) or sheets ($V \propto l$), the integrated
dissipation is therefore always dominated by the dissipation
which takes place on the smallest scales, because $p(V)
\epsilon_l \propto l^{-5}$ in the first case or $\propto
l^{-3}$ in the second case.

This confirms the important point for the evolution of
molecular clouds that dissipation of turbulence is
concentrated in a small subset of space. The induced
radiative cooling, and therefore the dissipation rate, have
to be searched on scales on the order of the milliparsec in
emission lines more powerful than the low$-J$ CO
transitions. The value of the rate itself may thus be
directly observable in line emissions (pure rotational lines
of \hh, \cp) that can only be distinguished from UV-excited
emission by observations at very high angular resolution.

\section{Conclusion and perspectives}

We performed a statistical analysis of the turbulence
towards two translucent molecular clouds based on the
two-point probability density functions of the \twCO\jone\
line centroid velocity.

Thanks to the excellent quality of the data, we prove the
non-Gaussian tails in the PDF of the line centroid velocity
increments on small-scales, down to a probability level of
\dix{-4}.  We show that the largest CVI, in both fields,
delineate elongated narrow structures ($\sim 0.02$~pc) that
are, in one case, parallel to the local direction of the
magnetic field.  In the Polaris field, these filaments are
well-correlated to the warm gas traced by the optically thin
\twCO\jone, while they do not follow the distribution of
matter traced by the \thCO. Using large-scale data, we have
shown that these filamentary structures remain coherent over
more than a parsec. Furthermore, the similar statistics
found in the IRAM and KOSMA maps of this field suggest that
both samples belong to the self-similar turbulent cascade.
In the Polaris field, the high-order structure function
exponents, computed up to order $p=6$, significantly depart
from their Kolmogorov value.

Through the properties of the tails of their PDF, \ie\ the
E-CVI, the line centroid velocities in these two clouds are
found to carry the main signatures of intermittency borne by
a turbulent velocity field. The departure from the Gaussian
statistics of the centroid velocity increments on
small-scales is therefore ascribed to the intermittency of
turbulence, \ie\ the non-space filling character of the
turbulent cascade.  The structures of largest CVI trace the
intermittent structures of intense shears and the sites of
intermittent turbulence dissipation. We show that these
intermittent structures, on the 0.02~pc-scale, harbour 25\%
and 5\% of the total energy dissipation, in the Polaris and
Taurus fields, respectively, although they fill less than
2.5\% and 1\% of the cloud area.  We find that both fields
are intermittent and that the more intermittent velocity
field on small scales (the Polaris field) belongs to a
molecular cloud far from virial balance on the scale of
30~pc. In contrast, the less intermittent (the Taurus field)
belongs to a virialized complex. The more turbulent field is
thus the more intermittent. Interestingly enough, the less
turbulent field is embedded in a star-forming cloud (Taurus
complex) with numerous young stellar objects, while the more
turbulent (Polaris) is in an inactive complex.

The exact nature of these intermittent structures, their
link with shocks, and the role of magnetic fields are still
elusive.  The comparison of observational data with
theoretical scalings requires the ability to compute higher
orders of the structure functions and establish the
correspondence between the centroid velocity and the
velocity fields. This stresses the need for large
homogeneous data samples with at least \dix{5} spectra. Such
data sets would also allow determination of the three
parameters of the class of models to which the SL94 or MHD
scalings belong.

Heterodyne instrumentation (\eg\ multi-beam receiver
heterodyne arrays) offers a dramatic increase in the spatial
dynamical range accessible, combining high spatial and
spectral resolutions. Sub-arcsecond resolution is needed to
resolve the dissipation scale, combined with a large
instantaneous field of view to disclose the shape of the
dissipative structures. Observational signatures of the
dissipation of the turbulent kinetic energy might be
searched for in chemical abundances of species, whose
formation requires high temperatures (Paper~I), like \chp,
\hcop, and water.  Excited \hh\ was also proposed as a good
coolant candidate \citep{falgarone2005,appleton2006}. While
some of these observational requirements are already met by
existing instruments (\eg\ HERA at the IRAM-30m telescope),
ALMA, SOFIA, and the Herschel satellite will definitely open
new perspectives in this field.
 
\acknowledgements{We thank the anonymous referee for his
  careful reading of the manuscript and useful comments that
  helped us to improve the paper. EF and PHB acknowledge the
  hospitality of the Kavli Institute for Theoretical Physics
  (Grant No. PHY05-51164) during the revision phase of their
  manuscript.  The authors also thank A.~Lazarian for useful
  comments and M.-A. Miville-Desch\^enes for providing them
  with the IRIS maps of Fig.~\ref{fig:polaris-iris}.}

\bibliographystyle{aa}


\begin{appendix}
\section{Two-point statistics}

\subsection{Construction of the \ncpdf}

\begin{figure}
  \begin{center}
    \includegraphics[width=0.7\hsize,angle=-90]{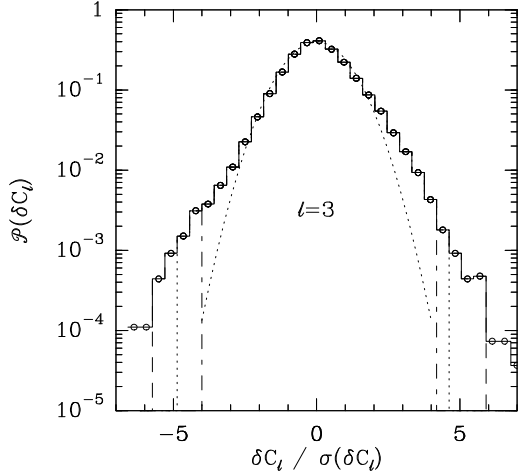}
	\caption{\ncpdf\ computed for $l=3$ in the Polaris
	  field, for successive values of \nmin=0, 10 (dashed),
	  30 (dotted), and 100 (dot-dashed).}
	\label{fig:nmin}
  \end{center}
\end{figure}

In each PDF, all the bins which are associated to a number
of events less than a given value \nmin, are blanked. The
value of \nmin\ depends on the number of bins in the
histogram. In Fig.~\ref{fig:nmin}, we show the \ncpdf\
computed for $l=3$ in the Polaris field, for successive
values of \nmin=0, 10, 30, and 100. It is seen that, with
\nmin=10, the spurious bins having a constant value
$\approx\dix{-4}$ are eliminated. The value of each bin and
its uncertainty are then determined from the average and rms
of all the points populating the bin.

\subsection{CVI maps}

\begin{figure}[tr!]
  \begin{center}
	\includegraphics[width=0.55\hsize,angle=-90]{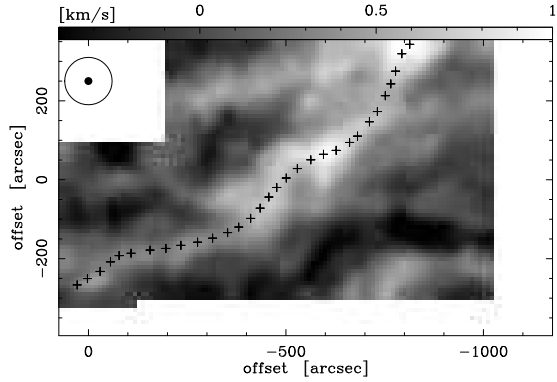}
	\caption{CVI map computed from the IRAM dataset with a
	  lag $l=18$ pixels (or 180\arcsec). CVI are in
	  \kms. The crosses indicate the positions used for the
	  cut of Fig.~\ref{fig:cvicut}. The KOSMA and IRAM beams
	  are indicated in the top left corner.}
	\label{fig:cvimap-lag19-pol}
  \end{center}
\end{figure}

The non-averaged CVI map of Fig.~\ref{fig:cvimap-lag19-pol},
computed in the IRAM data for a lag of 18 pixels (or
180\arcsec), shows that large-scale structures exist that
are not filtered out with large enough lags. The crosses
indicate the positions where the CVI values of
Fig.~\ref{fig:cvicut} have been taken.

\begin{figure}
  \begin{center}
    \includegraphics[width=\hsize]{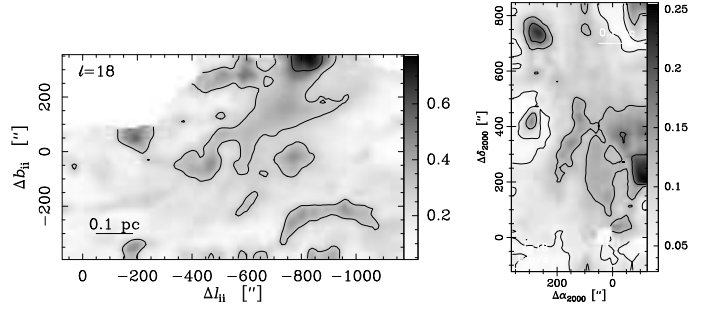}
    \caption{Same as Fig.~\ref{fig:cvimap-lag3} for a lag
      $l=18$ pixels. Weak patterns are visible, reminiscent
      of what is seen in
      Fig~\ref{fig:cvimap-lag3}. \textit{Left panel}:
      contours are the 0.29 and 0.58~\kms\
      levels. \textit{Right panel}: contours are the
      0.29~\kms\ with 0.29~\kms\ levels.}
    \label{fig:cvimap-lag18}
  \end{center}
\end{figure}

Figure~\ref{fig:cvimap-lag18} shows the CVI map computed in
the IRAM Polaris and Taurus fields, for a lag of 18
pixels. The comparison with the CVI maps of
Fig.~\ref{fig:cvimap-lag3} shows that the thin filaments
have faded away. However, in the Polaris map, the structure
visible at a lag of 3 pixels is still visible, though it has
broadened.

\subsection{Determination of the intermittency level}

\cite{she1994} developed a model to analyze the small-scale
properties of an incompressible turbulent flow.  Since this
model inspired numerous works of astrophysical relevance, we
summarize its key points here. SL94 propose studying the
large fluctuations of $\epsilon_l$ (defined as the
dissipation rate averaged over balls of size $l$) through
the ratio of its successive moments $\eeps{l}{p} =
\eps{l}{p+1} /\eps{l}{p}$. Hence, for each $p$, the value of
$\eeps{l}{p}$ describes the dissipation intensity of a set
of turbulent structures: as $p$ increases, the associated
structures are more coherent and more singular.  As a
result, the hierarchical structures of the SL94 model are
not related to any physical objects, except for the most
intermittent.  The SL94 model has three parameters: the
scaling of the velocity with scale $l$, assumed to be that
of the K41 theory ($\vcar{l} \sim l^{1/3}$); the level of
intermittency characterized by a parameter $\beta$; and the
dimensionality $D$ of the most intermittent structures.
Assuming $\beta=2/3$ and $D=1$, SL94 proposed a recursive
relation linking $\eeps{l}{p+1}$ to $\eeps{l}{p}$,
\begin{equation}
  \sfe{l}{p+1} = A_p {\sfe{l}{p}}^\beta {\sfe{l}{\infty}}^{1-\beta},
  \label{eq:recursive}
\end{equation}
which allowed them to compute the anomalous scaling of the
energy dissipation rate with scale $l$, $\eps{l}{p} \sim
l^{\tau_p}$, with $\tau_p = -2/3p + 2 [1 -
(\frac{2}{3})^p]$.

In principle, by fitting the exponents $\zz(p)$, it should
be possible to determine the three parameters of the SL94
class of models. In practice, a reliable determination of
the three parameters, \eg\ by least-square fitting the
exponent values, requires computing high-order structure
functions ($p>6$). With 6 orders, we could only determine
one parameter, the scaling of the velocity fluctuations in
the cascade $\vcar{l} \propto l^\theta$ and we found
$\theta=1/3$ which coincides with the K41 value.

\begin{figure}[h!]
  \begin{center}
	\def\wa{0.65\hsize}
    \includegraphics[width=\wa,angle=-90]{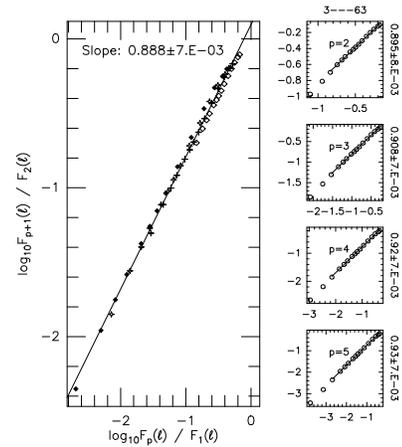}
    \caption{The $\alpha$-test to both the Polaris field
      (see Appendix~A.3) where $\alpha=\beta^\theta$ with
      $\theta=1/3$. All orders are plotted on the left panel
      and fitted with a single power law, while the right
      panels detail the fits to individual orders
      $p=2,3,4,5$. Fits are done over the range indicated by
      the line. The individual slopes are 0.90, 0.91, 0.92
      and 0.93.}
    \label{fig:btest}
  \end{center}
\end{figure}

\begin{figure}
  \begin{center}
	\def\wa{0.65\hsize}
	\includegraphics[width=\wa,angle=-90]{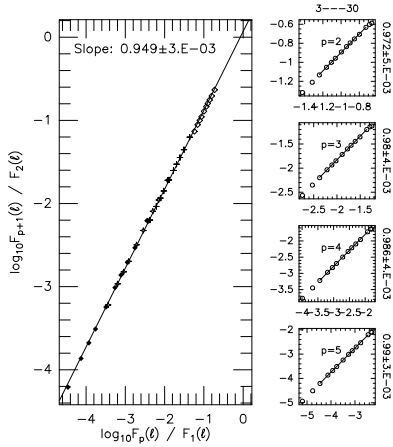}
    \caption{As Fig.~\ref{fig:btest} for the Taurus field,
      where the slopes for each order ($p=2,3,4,5$) are
      0.94, 0.95, 0.96 and 0.97.}
    \label{fig:btest2}
  \end{center}
\end{figure}

\cite{she2001} propose a method of determining the parameter
$\beta$ independently. It makes use of the fundamental
assumption of the hierarchical SL94 model that there is a
scaling law for the successive powers of the energy
dissipation on scale $l$. Using the Kolmogorov-Oboukhov
refined similarity hypothesis \citep[RSH, see
\eg][]{lesieur1997} $S_p\sim\eps{l}{p/3}l^{l/3}$, a
recursive relation similar to Eq.~\ref{eq:recursive} can be
written that involves functions of the ratio of successive
orders of the $S_p$:
\begin{equation}
  \sff{l}{p+1} = A_p {\sff{l}{p}}^\alpha
  {\sff{l}{\infty}}^{1-\alpha}
\end{equation}
with $\sff{l}{p}=S_{p+1}(l)/S_p(l)$. This recursive
relation, together with the RSH and the second assumption
that $\sff{l}{\infty}\sim S_3^\gamma$, leads to the
expression of the relative scaling exponents:
\begin{equation}
  \zz_p = p\gamma + (1-3\gamma)\frac{1-\alpha^p}{1-\alpha^3}
\end{equation}
with $\alpha=\beta^\theta$ and $\theta=1/3$. The
$\alpha$-test proposed by \cite{she2001} is to plot
$\sff{l}{p+1}/\sff{l}{2}$ against $\sff{l}{p}/\sff{l}{1}$
since $\sff{l}{p+1}/\sff{l}{2} = (A_p/A_1) \, \left(
\sff{l}{p}/\sff{l}{1} \right)^\alpha$.  If the log-log plot
is a line of slope $\alpha$, the data are said to pass the
$\alpha$-test. In Figs.~\ref{fig:btest}-\ref{fig:btest2}, we
show the result of the $\alpha$-tests applied to the Polaris
and Taurus data sets. Both pass the $\alpha$-test with
values of $\alpha=0.89\pm0.01$ and $0.950\pm0.005$ for the
Polaris and Taurus fields, respectively. The corresponding
values of $\beta$ are $0.70\pm0.04$ and $0.86\pm0.02$ in the
Polaris and Taurus fields, respectively (see
Figs.~\ref{fig:btest}-\ref{fig:btest2}). The intermittency
level in the Polaris field is very close to the SL94 value,
$\beta=2/3$, but our data sets are too small to allow a
determination of $D$.

\end{appendix}

\begin{appendix}
\section{Filling factor of the CVI and dissipation}

\begin{figure*}
  \begin{center}
	\def\wa{0.3\hsize}
	\includegraphics[width=\wa]{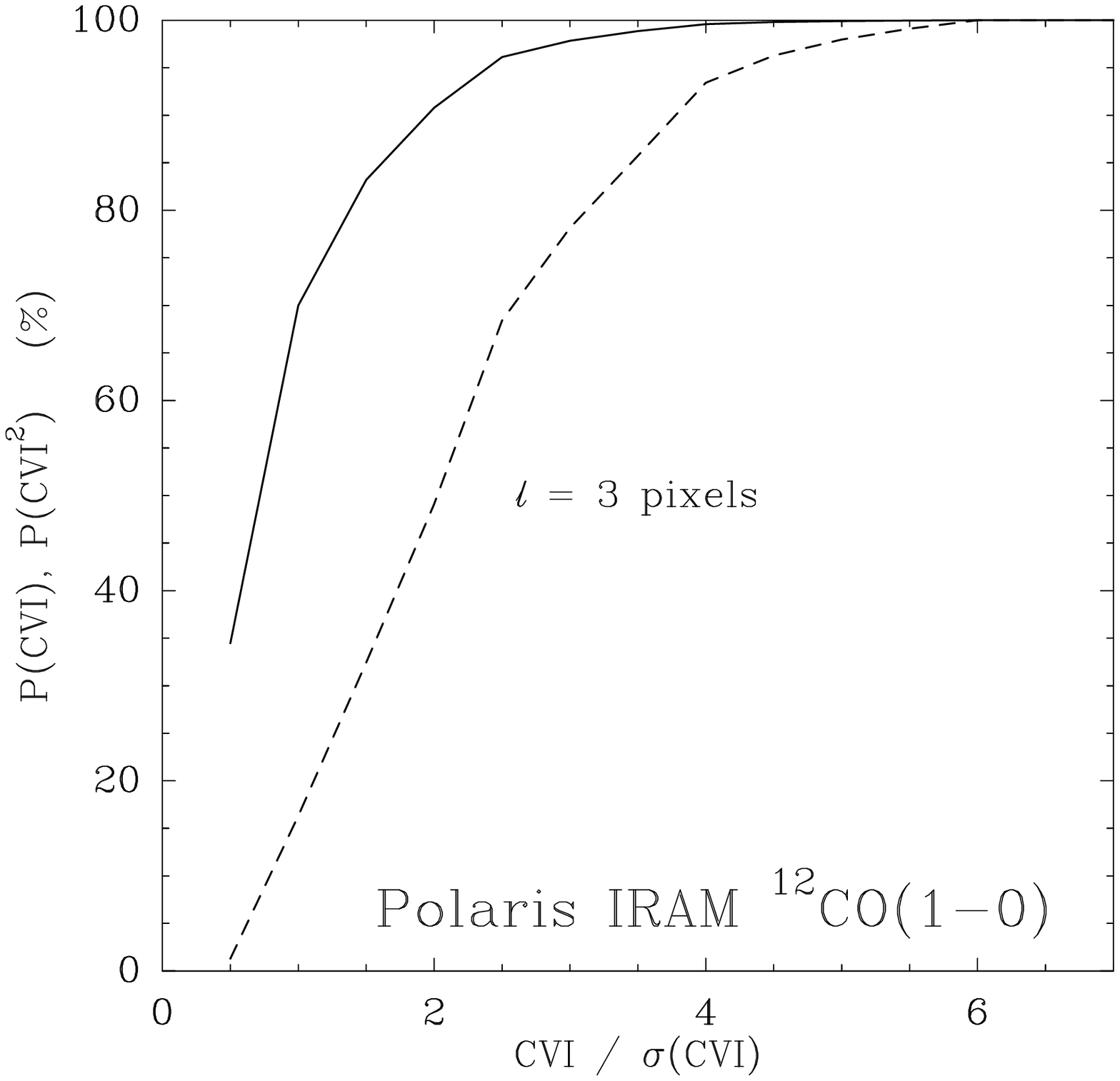}\hfill%
	\includegraphics[width=\wa]{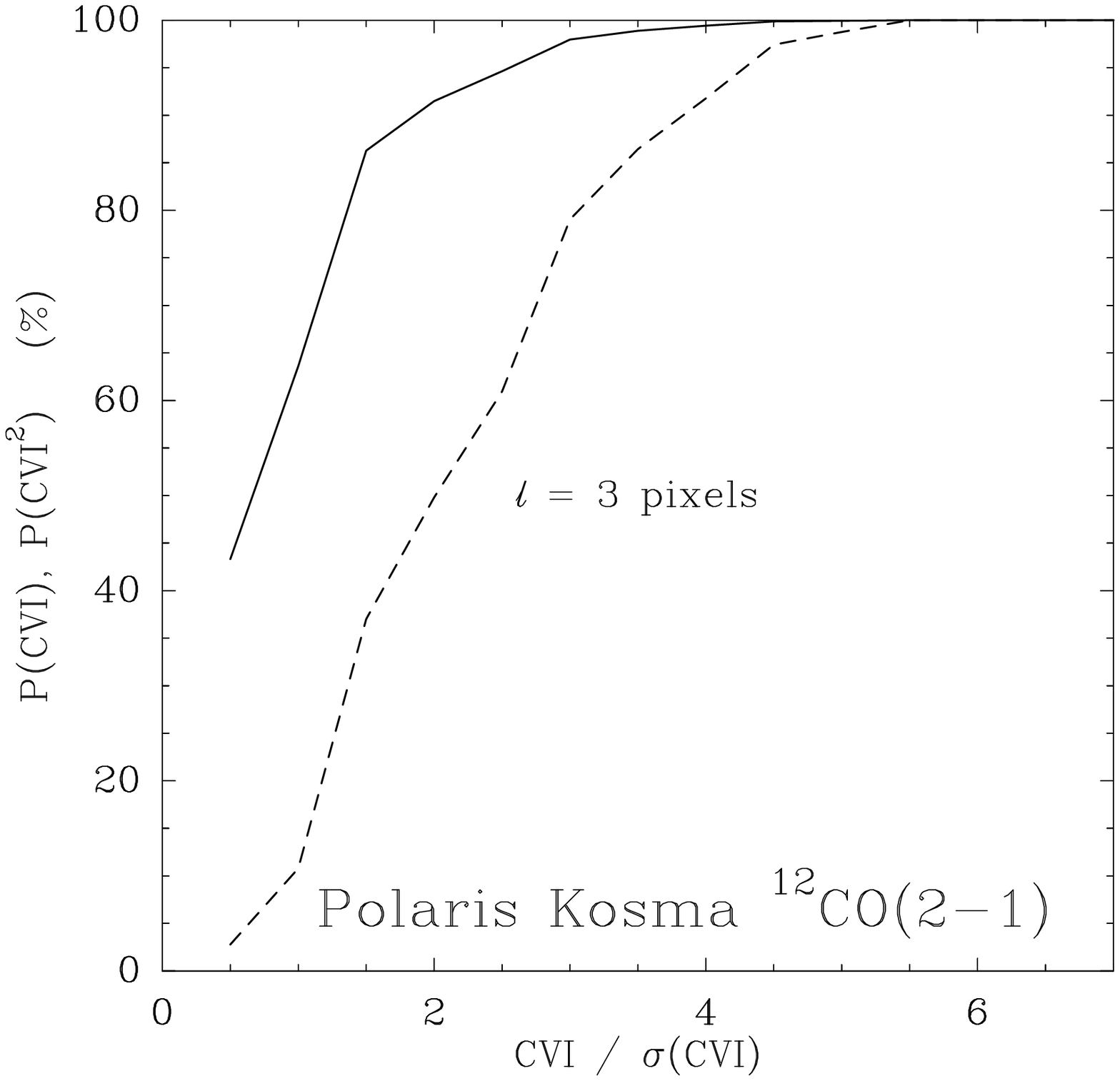}\hfill%
	\includegraphics[width=\wa]{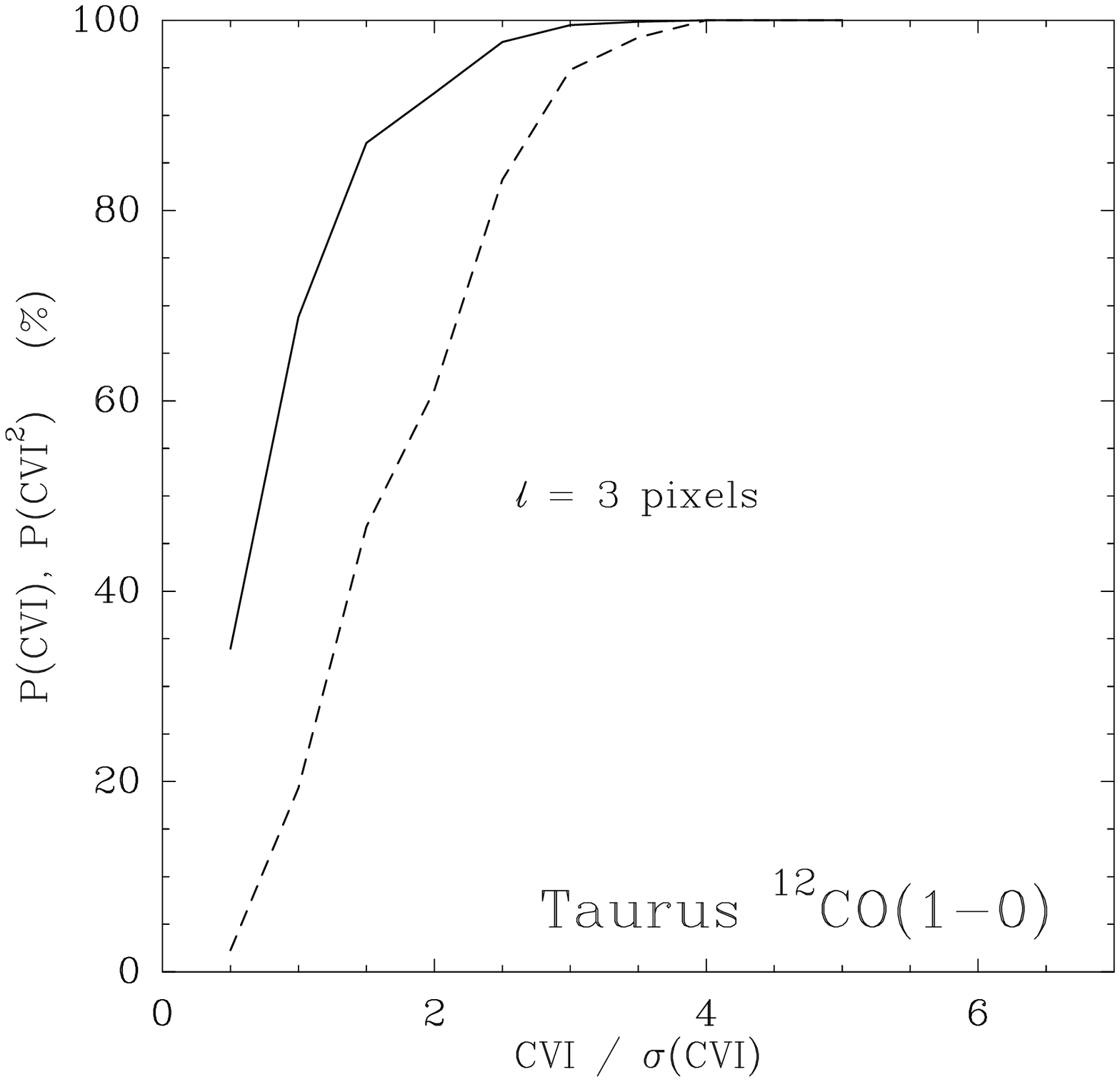}
	\caption{Cumulative distributions of the CVI (\dc, full
	  line) and the associated cumulative distributions of
	  $\dc^2$ (dashed line), shown as functions of the
	  normalized \dc\ (computed for a lag $l=3$).}
	\label{fig:pdf-dissip}
  \end{center}
\end{figure*}
Based on the PDF of CVI of Figs.~\ref{fig:pdf-polaris} and
\ref{fig:pdf-taurus}, we computed the cumulative fraction of
the CVI \dc\ and the associated cumulative fraction of
$\dc^2$. The non-averaged CVI \dc\ are more closely related
to the energy dissipation than the azimuthally averaged ones
(\avcvi) because they preserve the two derivatives of the
$los$ velocity. The results are shown in
Fig.~\ref{fig:pdf-dissip}, for a lag $l=3$. Assuming that
the dissipated energy scales as $\dc^2$, we see that in the
Polaris field, the regions with CVI larger than 3\sigdc\
represent only 2.5\% of the surface and they contribute to
$\approx 25\%$ of the energy dissipation. In the Taurus
field, 5\% of the energy dissipation is concentrated into
less than 1\% of the surface.

\end{appendix}

\end{document}